\documentclass[journal]{IEEEtran}
\IEEEoverridecommandlockouts
\usepackage{cite}
\usepackage{amsmath,amssymb,amsfonts}
\usepackage{algorithmic}
\usepackage{graphicx}
\usepackage{textcomp}
\usepackage{xcolor}
\usepackage{hyperref}
\usepackage{etoolbox}
\usepackage{cleveref}
\usepackage{subcaption}
\usepackage{textcomp, gensymb}
\usepackage{multirow}
\usepackage{tabularx,booktabs}
\usepackage{balance}
\newcolumntype{Y}{>{\centering\arraybackslash}X}
\robustify\,
\usepackage{siunitx}
\def\BibTeX{{\rm B\kern-.05em{\sc i\kern-.025em b}\kern-.08em
    T\kern-.1667em\lower.7ex\hbox{E}\kern-.125emX}}
\begin{document}

\title{Millimetre-wave Radar for Low-Cost 3D Imaging: A Performance Study}

\author{\IEEEauthorblockN{Han Cui, Jiacheng Wu and Naim Dahnoun}}

\maketitle

\begin{abstract}
Millimetre-wave (mmWave) radars can generate 3D point clouds to represent objects in the scene.
However, the accuracy and density of the generated point cloud can be lower than a laser sensor.
Although researchers have used mmWave radars for various applications, there are few quantitative evaluations on the quality of the point cloud generated by the radar and there is a lack of a standard on how this quality can be assessed.
This work aims to fill the gap in the literature.
A radar simulator is built to evaluate the most common data processing chains of 3D point cloud construction and to examine the capability of the mmWave radar as a 3D imaging sensor under various factors. 
It will be shown that the radar detection can be noisy and have an imbalance distribution.
To address the problem, a novel super-resolution point cloud construction (SRPC) algorithm is proposed to improve the spatial resolution of the point cloud and is shown to be able to produce a more natural point cloud and reduce outliers. 
\end{abstract}


\section{Introduction}
Millimetre-wave (mmWave) radars have received increased popularity in many industries as an emerging type of sensor.
The high bandwidth allows them to estimate the distance of an object at centimetre-level resolution, and the short wavelength and antenna size allow  multiple antennas to be integrated into a single chip and measure the angle-of-arrival (AoA) of the object using multiple-input multiple-output (MIMO) techniques \cite{mimo}. 
Combining the distance and AoA measurement, mmWave radars are able to construct a point cloud to represent the spatial shape of an object \cite{fundamental-ti}.
Therefore, mmWave radars can be used as a low-cost 3D imaging sensor, as an alternative to the traditional depth cameras and laser sensors.
This allows many computer vision tasks, such as object detection \cite{mmwave-object-detection}, human tracking \cite{mid,my-tworadars}, posture estimation \cite{mmpose,my-posture}, and identification \cite{mid}, to be addressed using mmWave radars as a non-intrusive solution.
However, although many applications have been proposed that rely on the 3D imaging capability of mmWave radars, few researchers have attempted to evaluate the quality of mmWave radars' detection quantitatively, and there is a lack of a standard on how this quality can be assessed.

When detecting objects in a scene, the reflection signal can be seen as a time-delayed version of the transmitted signal.
Combining the two signals gives an intermediate frequency (IF) signal, whose frequency and phase are determined by the time-of-flight (ToF) of the signal, and, equivalently, the distance between the object and the radar \cite{fundamental-ti}. 
The distance can be estimated directly from the IF signal of one pair of transmitter and receiver at high resolution, whereas the AoA needs to be estimated from the signal phase over a linearly spaced antenna array. 
There is rich literature on antenna array-based AoA estimation for traditional radars \cite{book-doa}, and the same concept also applies to mmWave radars.
This paper discusses the AoA estimation in the context of mmWave radar 3D imaging. 
It reviews and discusses the 3D point cloud construction techniques that are commonly used with mmWave radars. 

This paper presents a purpose-built simulation system that simulates the data acquisition process of a mmWave radar when facing a scene.
Radar data simulation allows researchers to focus on algorithm design and verification, instead of investing too much time in the hardware and real-world data collection. 
Existing radar simulators are often not designed for 3D imaging and have certain constraints. 
For example, the system in \cite{simulator-range} generates range and Doppler information of the radar rather than the raw data, the system in \cite{simulator-single} only supports single antenna data generation and cannot be used to estimate the AoA, and the system in \cite{simulator-complex} only supports up to four receivers in one direction and cannot be used for 3D imaging. 
In this research, a lightweight mmWave radar simulator is designed that supports raw data generation of a multi-antenna mmWave radar, configurable antenna parameters and layout, and customized scene construction using 3D human models with programmable motions. 

Using the simulation system, a quantitative evaluation of the radar's capability of imaging a human subject is carried out, as well as an evaluation of the key factors that could affect the output quality, including the data processing chain (DPC), radar antenna configuration, chirp configuration, subject velocity, and signal-to-noise ratio (SNR) in the scene. 
It will be shown that, although the radar can capture the spatial information of the subject's body shape, the detected point cloud can be noisy, sparse and imbalanced, and can require further processing before being used for higher-level applications.  
Finally, a novel super-resolution point cloud construction (SRPC) algorithm is proposed to improve the spatial resolution of the point cloud and is shown to be able to produce a more natural point cloud and reduce outliers. 

The contribution of this paper can be summarized as follows:
\begin{itemize}
    \item It presents a simulator of mmWave radar that can simulate the radar data as if it is placed in a real scene. It supports customized 3D models to be imported as the ground truth and provides a framework for evaluating a 3D imaging algorithm quantitatively.
    \item It presents a systematical study of 3D imaging algorithms using mmWave radars and an evaluation of the key factors that could affect the radar detection.
    It highlights the challenges of the noisy and imbalanced point cloud. 
    \item It presents a novel SRPC algorithm that can be inserted into the traditional point cloud construction DPC and can improve the quality of the point cloud. 
\end{itemize}

The rest of the paper is organized as follows. 
\Cref{sec:background} discusses the background and related work. 
\Cref{sec:preliminary} introduces the preliminaries of mmWave radars.
\Cref{sec:simulator} presents the details of the simulator.
\Cref{sec:pointcloud-algorithm} discusses the 3D imaging DPC using mmWave radars.
\Cref{sec:evaluation} presents the experimental results based on the simulation system. 
\Cref{sec:srpc} presents the novel SRPC algorithm and shows how it can improve the radar detection.
\Cref{sec:conclusion} concludes the work.

\section{Background}\label{sec:background}
Traditionally, 3D imaging systems often use depth cameras (like stereo cameras or RGBD cameras) \cite{review-depth-camera} or laser sensors \cite{review-lidar}, which are able to provide a dense and accurate 3D model of the object in front. 
However, camera-based systems can be intrusive and limited by the lighting conditions, and laser sensors are often constrained by their high cost (when compared with the cost of a mmWave radar which is only around \textsterling10). 
Radar-based 2D imaging has also been used widely in applications like security,
but they provide limited depth information and often rely on a dense antenna array that has a fixed region-of-interest \cite{review-mmwave-imaging-2d}. 
Radar-based 3D object detection uses radio frequency (RF) signals at certain frequencies to detect objects, and the resolution of the detection largely depends on the available bandwidth.
For example, WiFi devices operating at \qty{5}{GHz} with a \qty{40}{MHz} bandwidth can locate people with sub-meter level resolution \cite{wifi}, and ultra-wide band (UWB) devices at higher frequency bands can achieve centimetre level resolution \cite{uwb}. 
With mmWave radars operating at above \qty{60}{GHz}, the range resolution can be below \qty{5}{cm} \cite{fundamental-ti} and even micrometre level when pointing to a corner reflector \cite{mmwave-micrometre}.
Therefore, the high resolution has gained mmWave radars great popularity in automotive driving applications, and researchers are actively investigating their usage in computer vision tasks. 

Although mmWave radars can be used as 3D imaging sensors, the point cloud is often less accurate and noisier than the traditional systems \cite{my-tworadars}.
Many methods have been proposed to improve the detection quality of a mmWave radar, such as \cite{improve-pointcloud-dl,improve-pointcloud}. 
However, as these methods often use different radar configurations and scene setup, it is hard to carry out a quantitative comparison between them, and there lacks a standard on how to define the quality of the radar detection.
This work aims to address the problem by providing a simulation system and a framework for a systematic evaluation of a 3D imaging algorithm. 

Radar-based 3D imaging requires measuring the distance and AoA of the object.
The distance is often measured through the ToF of the signal, whereas the AoA measurement relies on the use of an antenna array. 
Since the antennas in the array will have different physical locations, the ToF at each antenna will be different, and the AoA of the object can be estimated by investigating the signal difference.
This process has been studied in depth for traditional long-range radars \cite{book-doa}, and the same principle can be applied to mmWave radars on a smaller scale. 
There are many algorithms designed for estimating the AoA based on a linearly spaced antenna array, such as the FFT-based method, beamforming method and subspace method (more details in \Cref{sec:angle}).
These algorithms provide a trade-off between the computational complexity and the angular resolution \cite{book-doa,aoa-compare}.
However, in contrast to traditional radar systems where the signal sources are often well-defined and uncorrelated, 
signal sources in 3D imaging can be one object with a continuous surface, which can require a different DPC.
This paper discusses the traditional AoA estimation algorithms in the context of 3D imaging using a mmWave radar and investigates the key factors that would affect the detection result. 

\section{mmWave Radar Preliminaries}\label{sec:preliminary}

Commercial mmWave radars often implement the frequency modulated continuous wave (FMCW) model. 
The radar sends a modulated chirp signal, detects the signal reflection from any object, 
processes the signal and determines the range, velocity, and AoA of the object. 
The principle of the FMCW radar model has been documented in detail in the literature (e.g. \cite{my-tworadars}). 
This section will give a brief discussion of the fundamentals that are necessary for understanding this paper, with a particular focus on the AoA estimation.

The radar sends an FMCW signal and receives its reflection from the object in the scene, where the reflection will be a time-delayed version of the transmitted signal.
The two signals are mixed to produce an IF signal, as shown in \Cref{eq:IF} (more details in \Cref{apx:if}):
\begin{equation}\label{eq:IF}
    IF(t) = A e^{j(\omega_{b} t + \phi_{b})} \ \mathrm{where} \ \omega_b = 2\pi S\tau, \ \phi_b = 2\pi f_0\tau
\end{equation}
where $S$ is the slope of the chirp, $\tau$ is the ToF of the signal, $f_0$ is the starting frequency of the chirp, and $A$ represents the amplitude of the signal. After obtaining the IF signal, a DPC will be applied to determine the presence of any object.

\subsection{Distance and Velocity Estimation}\label{sec:distance}
For a single object, the frequency $\omega_b$ will be a constant value and the distance of the object can be calculated as 
\begin{equation}\label{eq:distance} d = \frac{\tau c}{2}=\frac{\omega_b c}{4\pi S}\end{equation}
When there are multiple refection sources, the frequencies can be found by applying an FFT over the IF signal, which is referred to as the range-FFT. 
The velocity can be measured by transmitting multiple chirps at a known interval and calculating the phase difference between the chirps. 
Assuming the radar transmits a chirp every $T_c$ seconds and a phase difference $\Delta\phi$ is observed between successive chirps, then the velocity $v$ of the object can be estimated as:
\begin{equation}
    v = \frac{\Delta\phi c}{4\pi T_c f_0} \label{eq:velocity1}
\end{equation}
When there are multiple reflection sources moving at different velocities, they can be found by applying another FFT over the chirp phases, which is referred to as the Doppler-FFT.

\subsection{AoA Estimation Principle}
The AoA of the object can be estimated by having multiple antennas operating concurrently and by comparing the phase difference between neighbouring receivers. 
Due to the spatial location difference between the receivers, the signal received at each receiver will have a slight phase difference depending on the relative position of the receivers and the AoA.
The AoA can be computed in both azimuth and elevation directions, given that there exists more than one antenna in each direction.
The azimuth and elevation angles will be denoted as $\theta_a$ and $\theta_e$, respectively, or $\theta_{(a,e)}$ when referring to both of them. 

Assuming there are $N_a\times N_e$ linearly spaced receivers in the azimuth and elevation directions, and $M$ objects in different directions $\theta_{(a,e)m}$, then each object can be viewed as a signal source and the receiving antenna array will receive a signal (denoted as $x$) as a weighted sum of the $M$ data source:
\begin{equation}\label{eq:angle-problem}
    x^{(N_a\times N_e)} = \sum_{m=1}^M \alpha_m s(\theta_{(a,e)m})+n
\end{equation}
where $s(\theta_{(a,e)m})$ is the steering vector that represents the phase difference between receivers when a signal arrives with angle $\theta_{(a,e)m}$, $\alpha$ is an unknown parameter that models the signal transmission from the data source to the receivers, and $n$ is the noise.
The AoA estimation can be modelled as estimating the values of $\theta_{(a,e)}$ for each object $m$, given a set of receiver data ($x$).


\begin{figure}[t]
    \centering
    \includegraphics[width=0.8\linewidth]{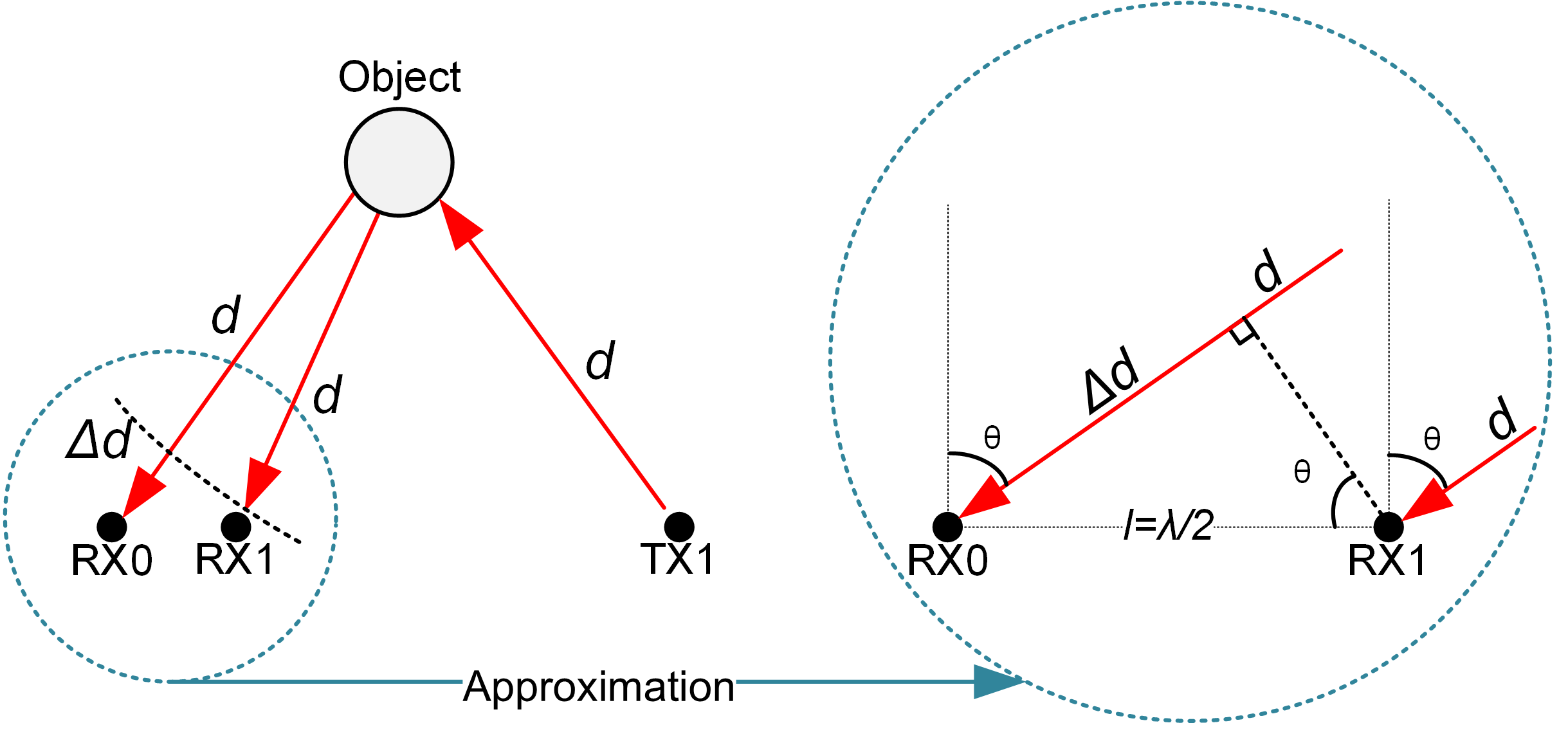}
    \caption{Phase difference between two receivers from one signal source.}
    \label{fig:angle}
\end{figure}

For linearly spaced arrays, the receivers are often separated by a small distance $l$ that is equal to half of the signal wavelength, i.e. $l=\frac{\lambda}{2}$, to maximize the angle-of-view (AoV) \cite{book-doa}.
When using an array of $N_a$ azimuth receivers and $N_e$ elevation receivers,
each subsequent receiver beyond the first one will receive an additional phase change that can be expressed using a 2D steering vector (more details in \Cref{apx:steering-vector}):
\begin{equation}\label{eq:steering-vector-2d}
\begin{split}
        & s(\theta_{(a,e)}, N_a, N_e) = \\
    &
    \begin{bmatrix}
    1                					 ,& ...,& e^{j(N_a-1)\Delta\phi_a}\\
e^{j\Delta\phi_e} ,& ...,& e^{j(\Delta\phi_e+(N_a-1)\Delta\phi_a)}\\
...,&...,&...\\
e^{j(N_e-1)\Delta\phi_e},& ...,& e^{j((N_e-1)\Delta\phi_e+(N_a-1)\Delta\phi_a)}
    \end{bmatrix}
\end{split}
\end{equation}

3D point cloud construction requires the x-y-z coordinates of the object instead of the azimuth and elevation angles.
Therefore, the calculation of the exact value of $\theta_a$ and $\theta_e$ is often not required. 
Let $d$ denote the distance of the object, then the 3D coordinates of the object can be calculated as (more details in \Cref{apx:steering-vector}):
\begin{equation}
x = d \frac{\Delta\phi_a}{\pi},\ z = d \frac{\Delta\phi_e}{\pi},\ y = \sqrt{d^2-x^2-z^2}
\end{equation}
Given that $d$ can be obtained from the range-FFT as discussed in \Cref{sec:distance}, the x-y-z coordinates can be obtained if the phase differences $\Delta\phi_a$ and $\Delta\phi_e$ are known.
Therefore, the AoA estimation of an object can be considered equivalently as searching for the best matching steering vector $s(\theta_{(a,e)m})$ of the object.

\subsection{AoA Estimation Algorithms}\label{sec:angle}
In the following sections, some of the most widely-used AoA estimation algorithms will be discussed, including the FFT-based method, conventional beamforming (also known as the Bartlett beamforming or the delay-and-sum beamforming), the minimum variance distortionless response (MVDR) beamforming (also known as the Capon beamforming) \cite{mvdr}, and the multiple signal classification (MUSIC) subspace method \cite{music}. 
The angle-FFT method is a single-snapshot method that can make an estimate based on a single chirp, whereas the other methods are multi-snapshot methods that require a few chirps to make one estimate. 
The performance of the algorithms depends on several factors, including the antenna layout, number of antennas, chirp configuration, number of snapshots, SNR, environment, etc. 

\subsubsection{Angle-FFT Method}\label{sec:angle-fft}
The simplest way of estimating $s(\theta_{(a,e)m})$ of an object $m$ in \Cref{eq:angle-problem} is by using correlation between the receiver data $x$ and the steering vector from the candidate angles.
A set of candidate steering vectors $s(\bar{\theta}_{(a,e)})$ is defined for $\theta_a \in [-\pi, \pi], \theta_e\in [-\pi, \pi]$, and the correlation is calculated as $s(\bar{\theta}_{(a,e)})\cdot x$, which will yield a peak output when $\bar{\theta}_{(a,e)}$ equals to $\theta_{(a,e)m}$. 
This process is equivalent to applying an FFT over the receiver data $x$, since the steering vector can be considered the same as a set of FFT coefficients, which gives the frequency components in terms of $\Delta\phi_a$ and $\Delta\phi_e$. 
This FFT is also referred to as the angle-FFT.

\begin{figure}[htbp]
\includegraphics[width=\linewidth]{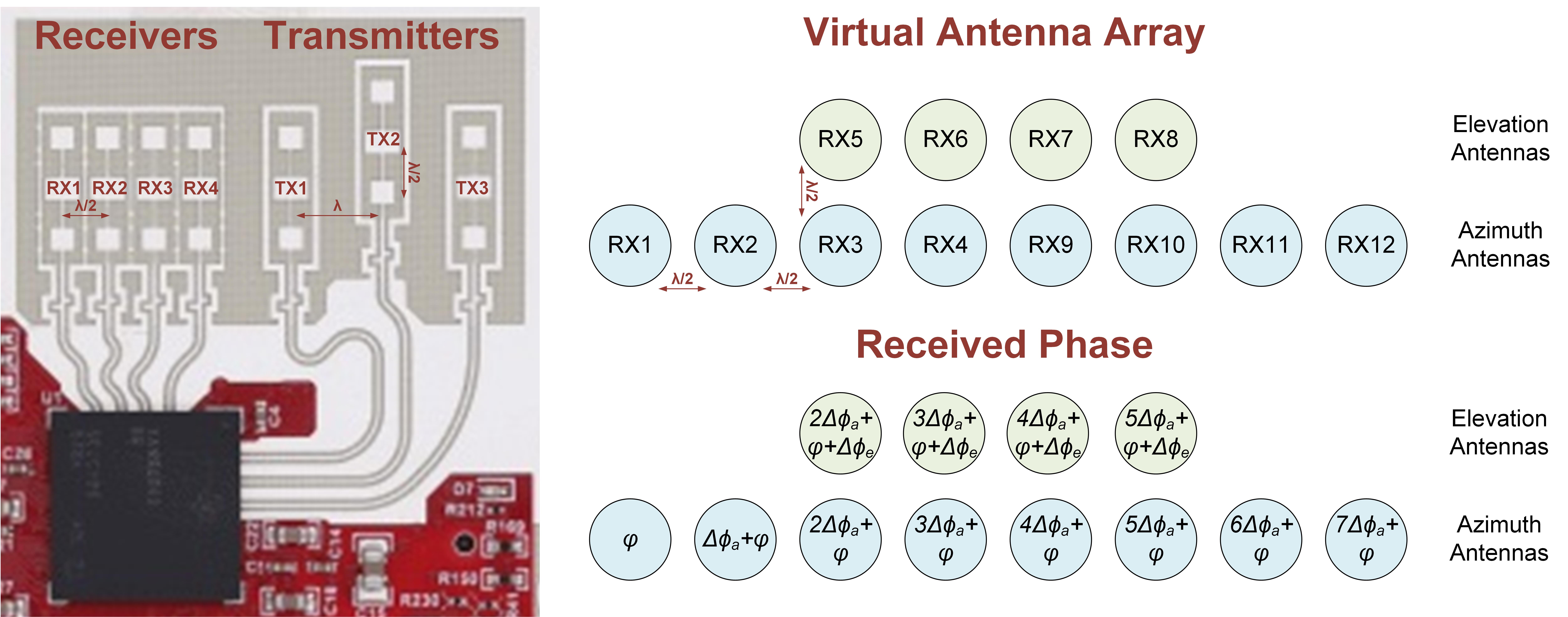}
\centering
\caption{IWR6843 radar antenna layout, the virtual receiver array and the received phases.}
\label{fig:antenna}
\end{figure}

As an example, \Cref{fig:antenna} shows the antenna layout of the TI IWR6843 radar. It has three transmitters and four receivers, which can form a 12-receiver array when using MIMO techniques \cite{mimo}. 
The phase of each virtual receiver is also shown, where $\varphi$ is the random initial phase of the first receiver. 
The azimuth receivers will form a signal $e^{j(\Delta\phi_a n+\varphi)}$ and the elevation receivers will form a signal $e^{j(\Delta\phi_a n+2\Delta\phi_a+\varphi+\Delta\phi_e)}$, where $n$ is the receiver index in each direction. 
The value of $\Delta\phi_a$ can be obtained by applying an azimuth-FFT over the azimuth receivers (RX1-RX4 and RX9-RX12), which will give the frequency $\Delta\phi_a$ and phase $\varphi$.
The value of $\Delta\phi_a$ can be obtained by applying an FFT over the elevation receivers, which will give the frequency $\Delta\phi_a$ and phase $2\Delta\phi_a+\varphi+\Delta\phi_e$.
Hence, the value of $\Delta\phi_e$ can also be calculated given $\Delta\phi_a$ and $\varphi$.

\begin{figure}[htbp]
\includegraphics[width=\linewidth]{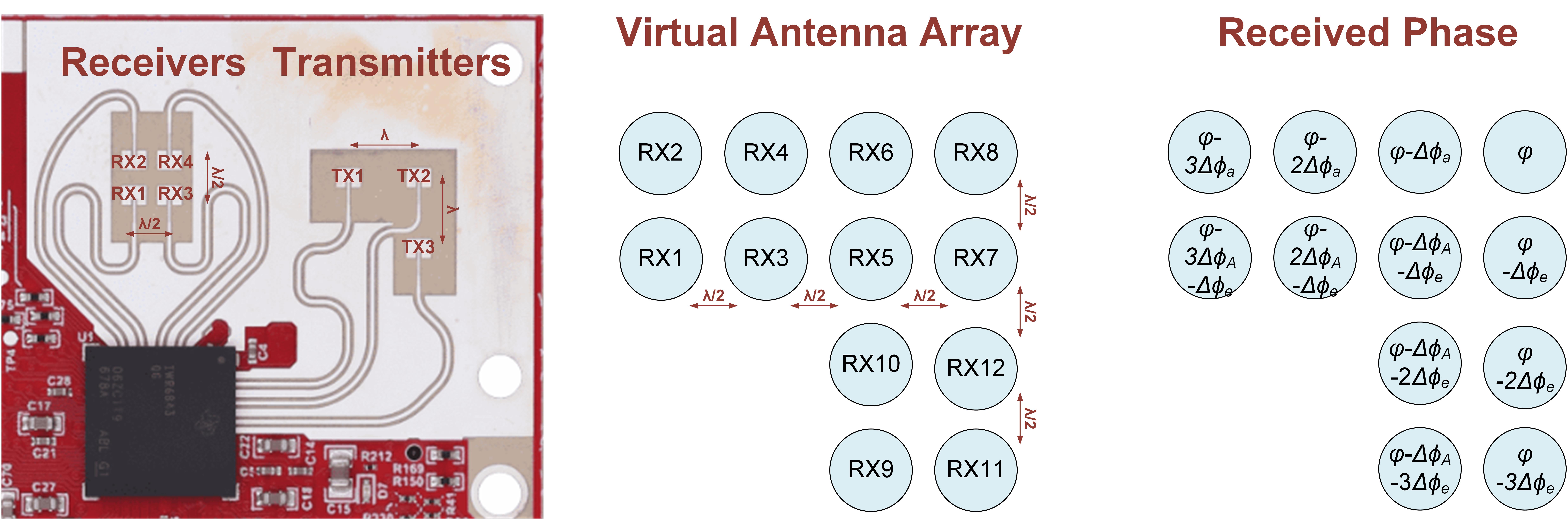}
\centering
\caption{IWR6843ODS radar antenna layout, the virtual receiver array and the received phases.}
\label{fig:antenna-ods}
\end{figure}

An alternative approach to calculate $\Delta\phi_a$ is by applying an elevation-FFT over a set of receivers in the elevation direction.
For example, \Cref{fig:antenna-ods} shows the layout of the TI overhead detection sensor (ODS) model, where the receivers form a near-square shape and allows a 2D angle-FFT to be performed. 
The ODS models allow a higher elevation resolution at the cost of reduced azimuth resolution.

\subsubsection{Beamforming Method}\label{sec:c2-bf}
Beamforming methods calculate a set of weights $w^{(N_{rx}\times \Theta)}$ for the $N_{rx}$ virtual receivers in the array (both azimuth and elevation), and for all possible angles $\theta_{(a,e)}\in \Theta$ where $\theta_a \in [-\pi, \pi], \theta_e\in [-\pi, \pi]$. 
When applying a column of weights to the receiver data $x$, the signal from the direction $\theta$ will receive a constructive inference.
By searching all possible angles $\theta_{(a,e)}$, a power spectrum $p$ with size $\Theta$ can be obtained, where a high power in the spectrum indicates that there is a data source in that direction:
\begin{equation}\label{eq:bf}
    p = w^H x
\end{equation}
where $w^H$ is the Hermitian transposition of $w$.
The angles of the $M$ objects can be obtained by taking the $M$ highest peaks in $p$ and finding the corresponding entries in $w$.

In the data model shown in \Cref{eq:angle-problem}, signals reflected from objects will be correlated when being received at each receiver, whereas the noise will be uncorrelated. 
Therefore, one way to extract signal information from $x$ is by calculating a sensor covariance matrix $R_x$ \cite{book-doa}:
\begin{equation}
    R_x = E\{x^H x\} \approx \frac{1}{N}\sum^N_{t=1} x^H(t) x(t)
\end{equation}
where $E$ represents the statistical expectation and $x(t)$ represents one snapshot (or one frame) of the receiver data $x$.
When evaluating the beamforming power spectrum using multiple snapshots, the overall power spectrum becomes the statistical expectation of $p$ in \Cref{eq:bf} over the snapshots, which gives:
\begin{equation}\label{eq:bf-cov}
    P = E\{|w^Hx|^2\} = \frac{1}{N}\sum^N_{t=1} w^H x(t)  x^H(t)  w = w^H  R_x  w
\end{equation}
Once the beamforming power spectrum is computed, the peaks in the spectrum will correspond to the signal from the objects.   

There are many algorithms designed for calculating the weights $w$.
The conventional beamforming uses the steering vector directly as the weights, which is conceptually equivalent to the angle-FFT method (or correlation-based method) in \Cref{sec:angle-fft}:
\begin{equation}\label{eq:conventional}
    P_{conventional} = s^H  R_x  s
\end{equation}
where $s$ is the candidate steering vector in the format of \Cref{eq:steering-vector-2d}.

There are also adaptive beamforming algorithms that calculate the weights using the signal information embedded in the covariance matrix.
For example, 
the MVDR algorithm aims at minimizing the variance from non-interested directions while keeping the signal from the candidate direction distortionless \cite{mvdr}:
\begin{equation}\label{eq:mvdr}
    P_{mvdr} = \frac{1}{s^HR_x^{-1}s}
\end{equation}

\subsubsection{Subspace Method}\label{sec:c2-subspace}
The core of the subspace method is that, since the signal $x$ should contain $M$ correlated signals and uncorrelated noise, the covariance matrix $R_x$ should have $M$ non-zero eigenvalues and $N-M$ zero eigenvalues, where $N$ is the rank of $R_x$ that is equal to the number of receivers. 
The eigenvectors corresponding to the $M$ eigenvalues form the signal subspace, and the eigenvectors corresponding to the zero eigenvalues form the noise subspace. 
The signal subspace and the noise subspace are orthogonal. 
One of the most widely-used subspace-based algorithms is the MUSIC algorithm \cite{music}. It searches for steering vectors that are orthogonal to the noise subspace. 
The power spectrum of the MUSIC algorithm can be written as:
\begin{equation}\label{eq:music}
    P_{music} = \frac{1}{s^HUU^Hs}
\end{equation}
where $U$ is the set of eigenvectors corresponding to the zero eigenvalues.

\section{mmWave Radar Simulator}\label{sec:simulator}
A simulator is designed to verify the discussed algorithms and evaluate the theoretical capability of using a mmWave radar as a 3D sensor. 
The simulator simulates mmWave radars with one transmitter and one receiving antenna array, which is practically equivalent to a multi-transmitter multi-receiver radar using an appropriate modulation scheme \cite{mimo}. 
Any two neighbouring receivers in the array are separated by $\lambda_0/2$, where $\lambda_0$ (approximately \qty{3.9}{mm}) is the wavelength of the mmWave signal at its chirp starting frequency (\qty{77}{GHz}).

The simulator simulates the IF signal at each receiver of a mmWave radar when pointing toward a scene. 
The scene is modelled to have $M$ points, where each point has a unique x-y-z coordinate and represents the spatial location of the object in the scene. 
Each point is modelled as a corner reflector and reflects the mmWave signal sent out by the radar with the same reflectivity.
The reflection area of the object is modelled by the number of points, i.e. a large object would have a higher number of points. 
The IF signal at a receiver during one chirp is modelled using \Cref{eq:IF}.
Given a certain chirp configuration, the frequency and phase of the IF signal from one point are determined by the distance $d$ between the point and the receiver. 
The amplitude of the IF signal is set to be inversely proportional to $d^4$, to simulate the power loss due to distances according to the radar range equation \cite{radar-range-equation}.
The final IF signal at a receiver is the accumulated IF signals from all $M$ points in the scene, with an additional white Gaussian noise $n$, as shown in \Cref{eq:if-simulation}.
\begin{equation}\label{eq:if-simulation}
    IF(t) = \sum_{i=1}^M \frac{1}{d_i^4}e^{j(2\pi S\tau_i \cdot t+2\pi f_0 \tau_i)} + n
\end{equation}
where $\tau_i$ is the ToF of the signal from the transmitter to the point $i$ and then to the receiver, and $S$ is the slope of the chirp. 
The amplitude of the noise $n$ is controlled by the desired SNR during the experiment.  
The signal $IF(t)$ is sampled into a digital signal of length $N_s$, where $N_s = \text{(duration of the chirp)} \times \text{(ADC sampling rate)}$.
During one chirp, the radar receives a signal that can be represented as a 2D matrix of size $N_{rx}\times N_s$, where $N_{rx}$ is the number of receivers in the array.
One frame includes $N_c$ chirps that form a 3D matrix of size $N_{rx}\times N_c \times N_s$, which becomes the input matrix of the point cloud construction algorithm, as shown as the input block in \Cref{fig:pointcloud-dpc}.

The design of the simulation system makes two assumptions. 
First, the multipath effect is not considered in this system. 
 While the multipath effect is a long-standing issue that can cause power fading and ghost targets, it highly depends on the scene and the reflectivity of the objects and is hard to incorporate in the model, so it is left as future work.
 Second, a practical radar often uses multiple transmitters and receivers and an appropriate signal modulation scheme to separate the signal from different transmitters, such as time demultiplexing modulation and binary phase modulation \cite{mimo}, to achieve an equivalent single-transmitter-multi-receiver system. 
 The simulation system assumes a perfect signal modulation scheme for this purpose and ignores any error or SNR loss that may be introduced during the modulation process.

\section{Point Cloud Construction Algorithm}\label{sec:pointcloud-algorithm}
The construction of a point cloud takes an input matrix of size $N_{rx}\times N_c \times N_s$ and outputs a 2D matrix $PC_K$ of size $K\times 3$ (referred to as the output point cloud), where $K$ is the number of detected points and $3$ is the x-y-z coordinates. 
This section studies one of the most common DPCs used on mmWave radars and its variant, which have shown success in many HAR systems, like in \cite{mid,mmpose,mmwave-har-activity-classification}.

\subsection{Data Processing Chains}\label{sec:c3-dpc}
\begin{figure}
    \centering
    \includegraphics[width=\linewidth]{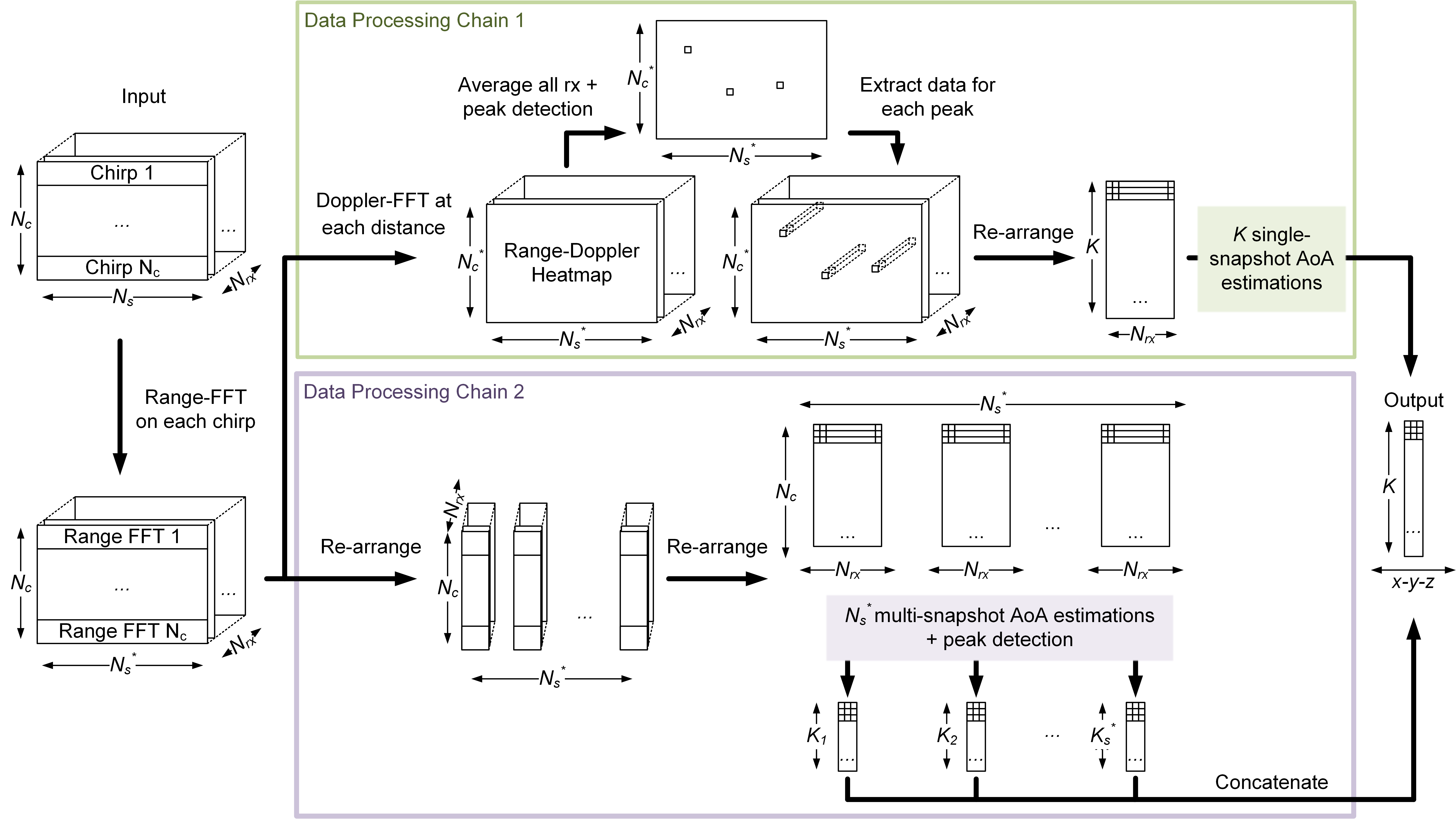}
    \caption{Two possible DPCs for mmWave radar point cloud construction.}
    \label{fig:pointcloud-dpc}
\end{figure}
Two DPCs are implemented that differ in using a Doppler-FFT or not, as shown in \Cref{fig:pointcloud-dpc}.
Both DPCs require a range-FFT over the raw data.
The range-FFT identifies the frequency components in the IF signal that correspond to the distance of an object.
It transforms the input matrix $X$ of size $N_{rx}\times N_c \times N_s$ into a range matrix $R$ of size $N_{rx}\times N_c \times N_s^*$, where $N_s^*$ is the length of the range-FFT.
The first DPC applies a Doppler-FFT on the data from all the chirps and generates a Range-Doppler heatmap of size $N_{rx}\times N_c^* \times N_s^*$, where $N_c^*$ is the length of the Doppler-FFT. 
Then, it searches for peaks in the Range-Doppler heatmap (using the average of all receivers), extracts the receivers' data for each peak and generates a 2D matrix of size $K\times N_{rx}$, where $K$ is the number of detected peaks and, equivalently, the number of detected points. 
A constant false alarm rate (CFAR) algorithm is used for detecting peaks from the Range-Doppler heatmap.
The parameters of the CFAR control the sensitivity of the peak detection and are considered the hyperparameters of the system. 
Finally, a single-snapshot AoA estimation is applied to each point in the matrix for a total of $K$ times, to obtain the x-y-z coordinates of all detected points. 
The AoA estimation algorithm can be any of the angle-FFT, beamforming or subspace methods. 
Although the beamforming and subspace methods are multi-snapshot algorithms, the Doppler-FFT implicitly uses the information from all chirps and allows a good estimate of the covariance matrix at the AoA estimation stage.

The second DPC does not include a Doppler-FFT.
Instead, it considers the chirps as different snapshots and performs one multi-snapshot estimation for each range bin for a total of $N_s^*$ times. 
More specifically, the input range matrix of size $N_{rx}\times N_c \times N_s^*$ is re-arranged into $N_s^*$ instances of $N_{rx}\times N_c$ matrix, and the AoA estimation is applied to each $N_{rx}\times N_c$ matrix using $N_c$ snapshots. 
The AoA estimation algorithm can be any of the beamforming or subspace methods. 
Finally, the points detected at each range bin are concatenated into one point cloud. 
In this research, the angle-FFT, conventional beamforming, MVDR beamforming and MUSIC subspace methods described in \Cref{sec:angle} are being studied. 

\subsection{Model Order Estimation}
As described in \Cref{sec:c2-bf} and \Cref{sec:c2-subspace}, the beamforming and subspace methods include an angle power spectrum computation step, where each peak in the spectrum corresponds to an incoming signal from a point. 
However, in both DPCs, the expected number of incoming signals will be unknown in practice.
Therefore, this number needs to be estimated from the signal data.
This step is referred to as model order estimation.
For this purpose, the covariance matrix of the signal data and its eigenvalues are computed.
As described in \Cref{sec:c2-subspace}, the covariance matrix should have a size of $N_{rx}\times N_{rx}$ and has a full rank equal to $N_{rx}$. There should be $M$ large eigenvalues that correspond to the number of incoming signals and $N_{rx}-M$ zeros corresponding to noise.
In practice, due to the presence of noise, the difference between these eigenvalues may not be significant. 
Therefore, the minimum descriptive length (MDL) algorithm \cite{mdl} is used for estimating the value of $M$. 
It fits a statistical model using the eigenvalues and searches for the optimal value of $M$ that minimizes a cost function. 
The MDL algorithm is used in both DPCs to estimate the number of incoming signals in the AoA estimation stage.
Once the angle power spectrum is calculated, all the local maxima will be found and the largest $M_{mdl}$ peaks will be taken as the output, where $M_{mdl}$ is the value found from the MDL algorithm. 

\subsection{Steering Vector Searching}\label{sec:2d-bf}
\begin{figure}
    \centering
    \begin{subfigure}[t]{0.32\linewidth}
        \includegraphics[width=\linewidth]{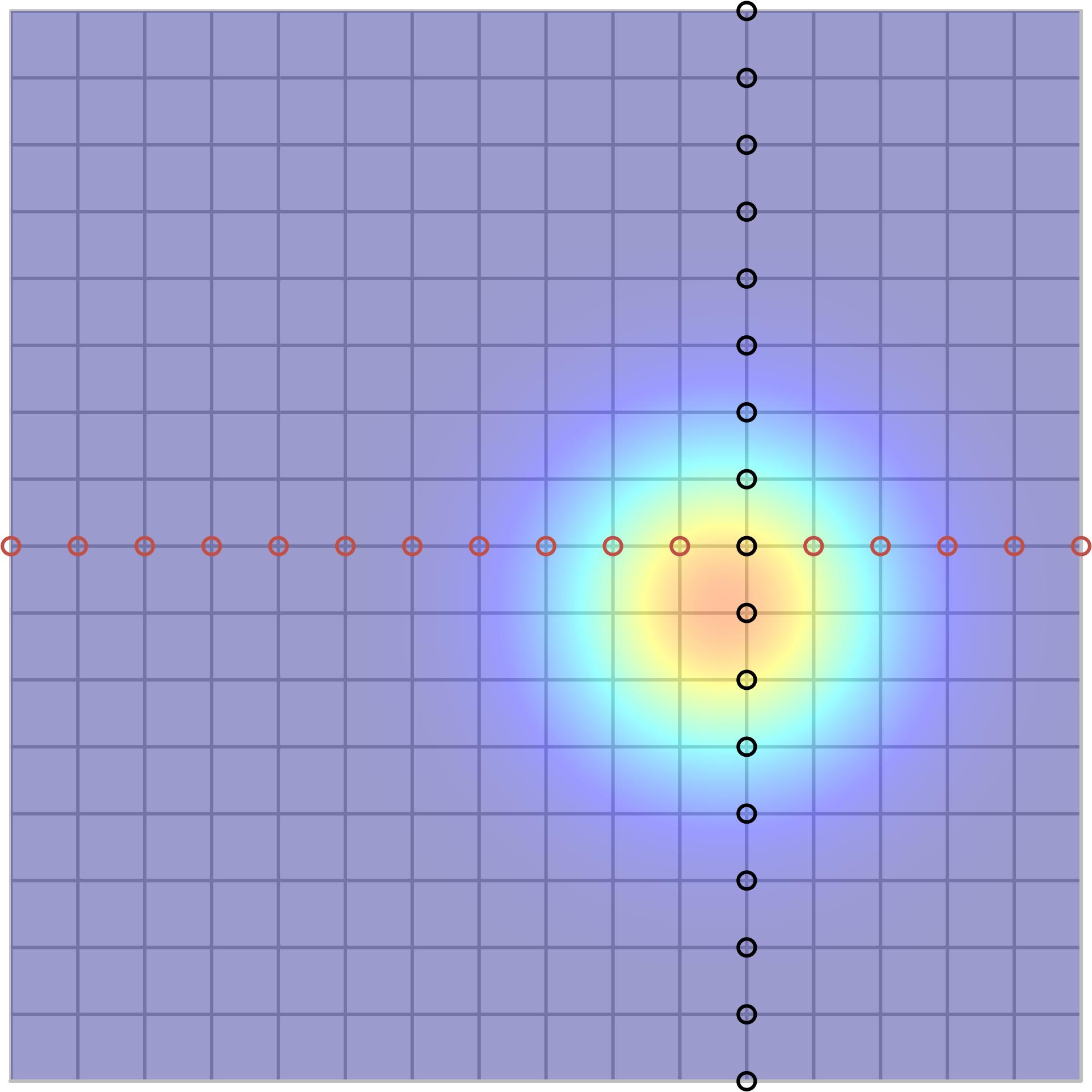}
        \caption{}
    \end{subfigure}
    \begin{subfigure}[t]{0.32\linewidth}
        \includegraphics[width=\linewidth]{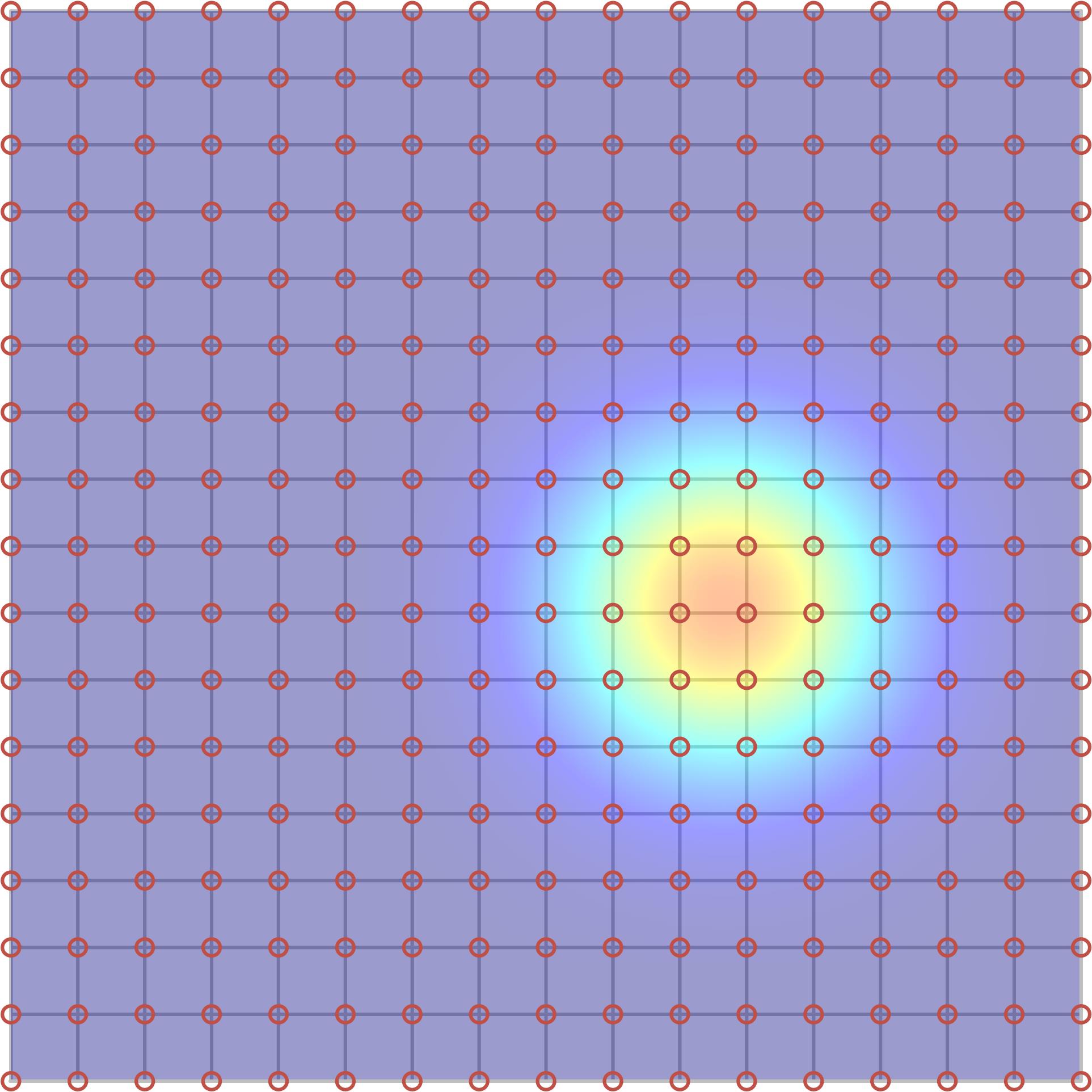}
        \caption{}
    \end{subfigure}
    \begin{subfigure}[t]{0.32\linewidth}
        \includegraphics[width=\linewidth]{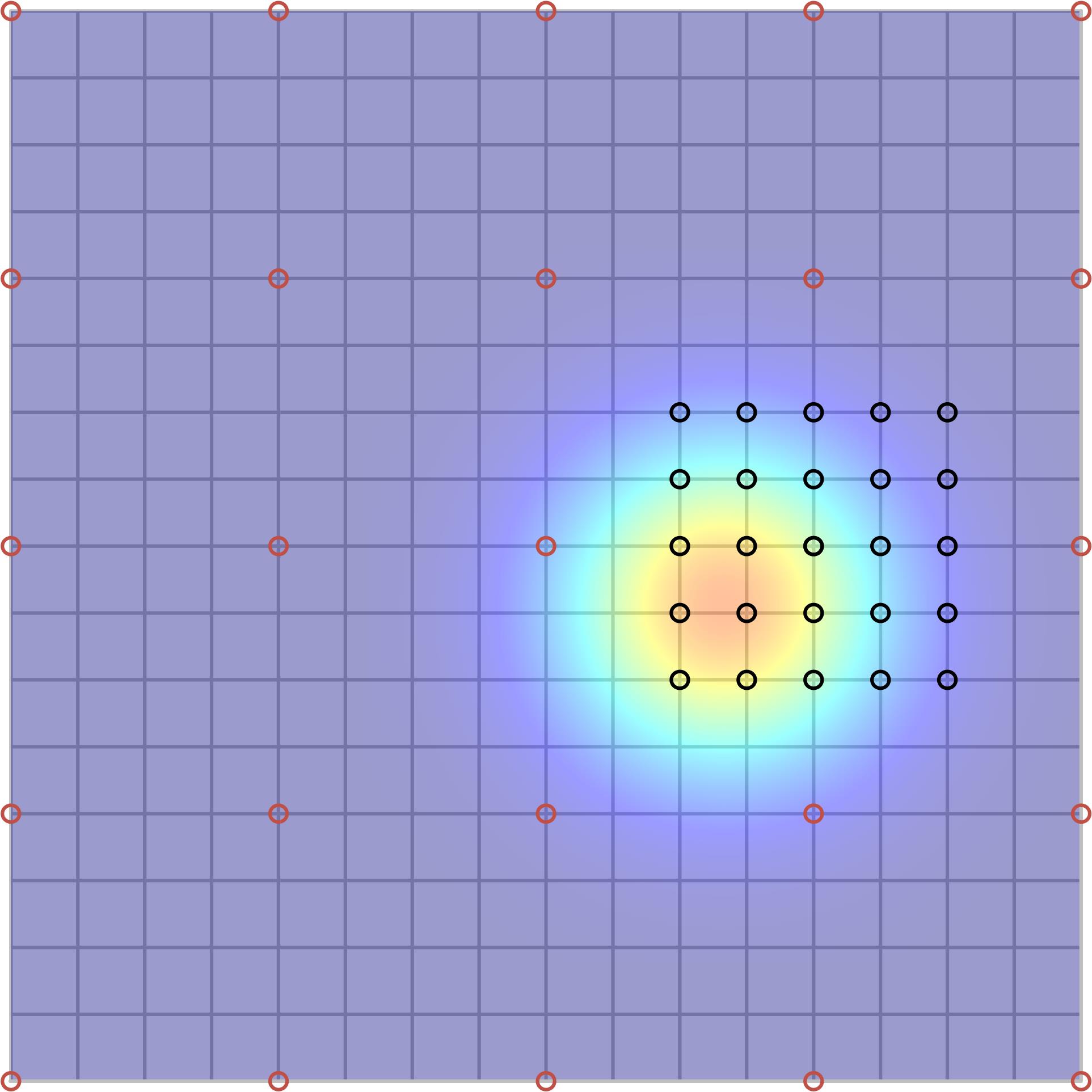}
        \caption{}
    \end{subfigure}    
    \caption{Three approaches when searching for the steering vectors. (a) An azimuth search (red) followed by an elevation search (black). (b) A full 2D azimuth-elevation search. (c) A 2D azimuth-elevation search using sub-grids.}
    \label{fig:2d-bf}
\end{figure}

The beamforming and subspace methods search for the steering vectors that maximize a power function. 
This process can be carried out using three approaches: an azimuth search followed by an elevation search, a 2D azimuth-elevation search or a 2D search using sub-grids.
An example of the three approaches is shown in \Cref{fig:2d-bf}.
In the example, the power spectrum shows the incoming direction of the signal.
The space of the spectrum is sampled into a $17\times17$ grid and each vertex on the grid represents a candidate AoA to be tested. 
In the first approach, an azimuth AoA search is performed using the data from azimuth receivers and steering vectors that only consider the azimuth angle.
Then, based on the azimuth AoA output, a secondary search is performed in the elevation direction using the data from all receivers.
This approach has the least computational cost (34 searches in the example), but the performance can be suboptimal as the azimuth search may not cover the actual AoA.
The second approach performs a 2D search that considers all possible combinations of the azimuth and elevation directions and uses data from all receivers.
It is computationally expensive (289 searches in the example) but provides the most accurate estimate.
The third approach defines several levels of grids and performs the AoA search at different granularities. 
It starts the searching with a sparse grid, finds the peaks, defines a denser grid around each peak and performs the next search. 
The process can be performed iteratively until the desired resolution is achieved.
It reduces the computational cost of the second approach significantly as it skips certain regions in the spectrum (50 searches in the example), at the cost of the potential possibility of missing some peaks.

\section{Evaluation}\label{sec:evaluation}
\subsection{Dataset}
The FAUST dataset \cite{faust} 
was used to serve as the ground truth for the simulator, to evaluate the point cloud construction algorithms described.
The datasets contain human models in the form of watertight triangulated meshes. 
The meshes were generated from a high-resolution camera system containing stereo cameras, RGB cameras and speckle projectors. 
The FAUST dataset contains 10 subjects and 30 static postures per subject, of which 10 postures are provided with aligned watertight models, giving 100 models in total. 

\begin{figure}
    \centering
    \includegraphics[width=\linewidth]{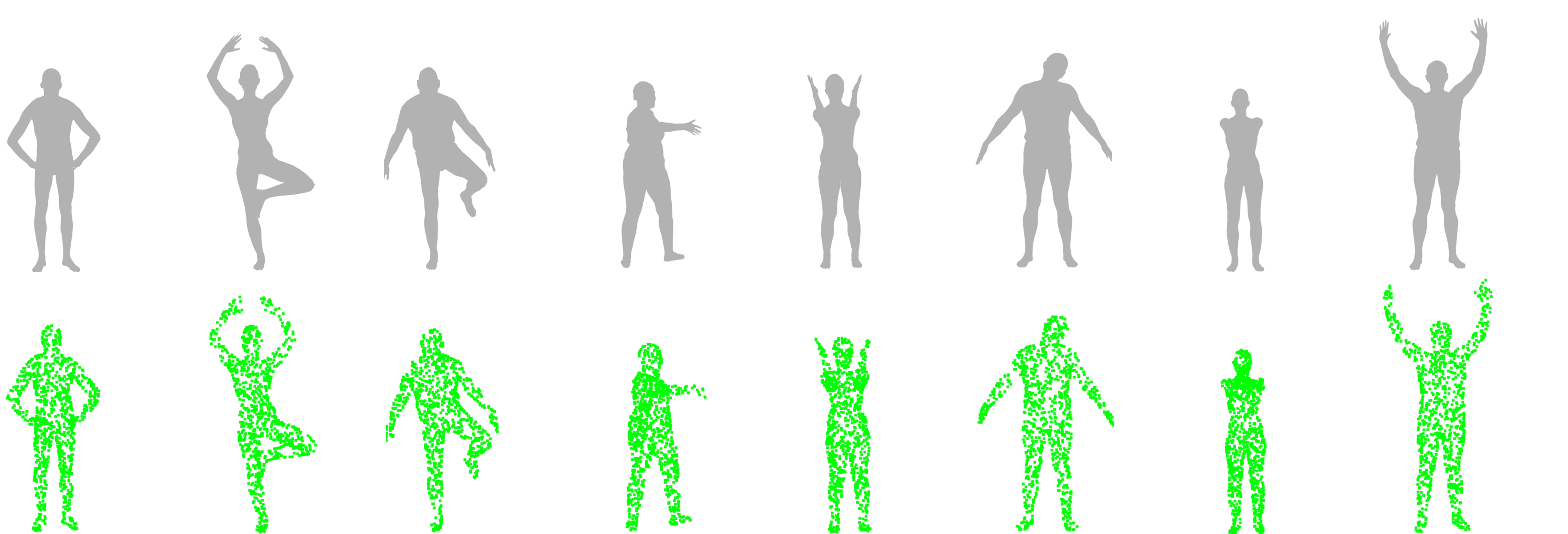}
    \caption{Some examples of the mesh models and point clouds from the FAUST dataset.}
    \label{fig:faust}
\end{figure}

In the simulation, the models were placed at \qty{2}{m} from the radar and facing towards the radar. 
The height of the radar was set to be in the middle of each model.
A ground truth point cloud was constructed from each model by randomly sampling $M$ points from the surface of the mesh model, where each point was assumed to be a corner reflector. 
Some examples of the mesh models and point clouds are shown in \Cref{fig:faust}.
The simulator computed a signal matrix for each point cloud to simulate the IF signal that would be received by the radar when placed towards a subject, as described by \Cref{eq:if-simulation}. 
The entire dataset containing the 100 models was split into 80 training data and 20 test data, where the training data was used for hyperparameters searching in the point cloud construction algorithms, and the test data was used for evaluating the algorithms. 

When generating the IF signal matrix, there are two sources of randomness: the noise term $n$ introduced in \Cref{eq:if-simulation} and the random sampling of the ground truth point cloud from the mesh model. 
Therefore, all the evaluation processes were repeated 10 times for each mesh model and the average metrics were reported, to minimize any potential effect of the randomness.

\subsection{Evaluation Metrics}
To evaluate the quality of the point cloud constructed by an algorithm, it is necessary to define the evaluation metrics for comparing the output point cloud against the ground truth point cloud. 
Let $PC_M$ denote the ground truth point cloud and $PC_K$ denote the point cloud generated by the radar, which are a $M\times 3$ matrix and a $K\times 3$ matrix, respectively. 
It is important to note that the point cloud construction algorithm can provide an uncertain number of points that might be different to the ground truth ($M\neq K$), and $PC_K$ can have a non-uniform distribution while $PC_M$ is distributed uniformly on the mesh model. 
The evaluation metrics should take the two point clouds $PC_M$ and $PC_K$ as input and measure the similarity between them. 
First, two points are defined to be close to each other if their Euclidean distance is less than a certain distance $D$.
In this research, $D$ is set to \qty{10}{cm} as an empirical estimation of the error tolerance of a HAR system. 
Then, the following terms and metrics are defined:
\begin{itemize}
    \item Precision: Number of points in $PC_K$ that has at least one close point from $PC_M$, divided by $K$. 
    It evaluates how many points in $PC_K$ are considered to be accurate.  
    \item Sensitivity/Recall: Number of points in $PC_M$ that has at least one close point from $PC_K$, divided by $M$. It evaluates how well $PC_K$ can cover the space of $PC_M$. 
    \item Fowlkes–Mallows index (FMI): the geometric mean of precision and sensitivity, i.e.
    $\sqrt{\text{precision}\times \text{sensitivity}}$.
    \item Intersection over Union (IoU): Establish two regular 3D voxel grids for $PC_K$ and $PC_M$ with the voxel size set to \qtyproduct{10x10x10}{cm}, consider a voxel to be occupied if there is at least one point present in the voxel, then the IoU is calculated as the number of overlapping voxels of the two voxel grid, divided by the union. 
    The IoU evaluates the similarity of the two point clouds at the granularity of the voxel size. 
\end{itemize}
An ideal system should have both high precision and high sensitivity, whereas the relative importance of the two depends on the application. 
In this section, the FMI, i.e. the geometric mean of precision and sensitivity, is used to indicate the performance of the system.
The IoU also provides a good indication of how the generated point cloud can represent the scene. 
However, as the calculation of the IoU is highly sensitive to the voxel size and outliers, it is used as a secondary metric. 

\subsection{Data Processing Chain and Algorithms}\label{sec:pc-baseline}
\begin{figure}
    \centering
    \includegraphics[width=0.9\linewidth]{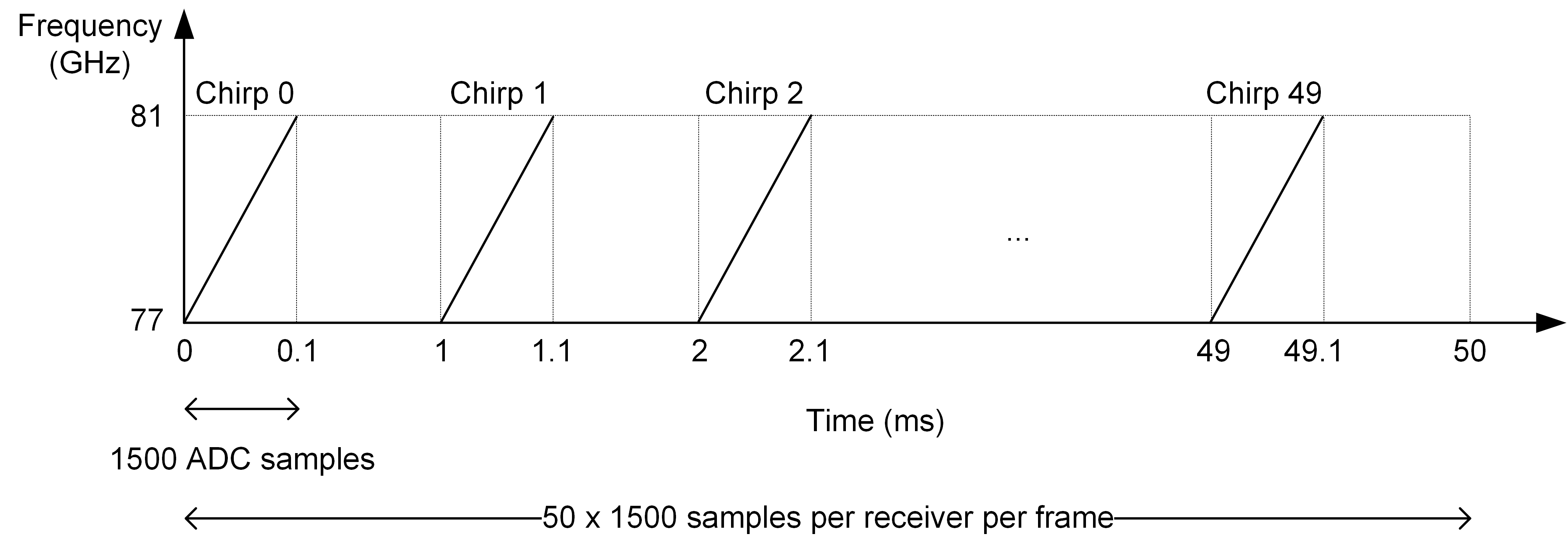}
    \caption{Chirp configuration of one frame in the baseline setup.}
    \label{fig:chirpcfg}
\end{figure}

In the first experiment, the two DPCs combined with different AoA algorithms were evaluated and compared, in terms of the quality of the estimated point cloud and the computational cost.
A baseline radar and scene configuration were designed to approximate a typical setup in a common indoor environment as follows:
\begin{itemize}
    \item One transmitter and a $4\times 4$ uniform receiver array.
    \item The chirp frequency is \qtyrange{77}{81}{GHz}, the slope is \qty{40}{MHz/us}, the chirp duration is \qty{100}{us}, the ADC sampling rate is \qty{15}{MHz}, each frame is \qty{50}{ms} with 50 chirps, and each chirp has 1500 samples (as shown in \Cref{fig:chirpcfg}). 
    \item Each human mesh model is sampled into 512 points and placed at \qty{2}{m} away from the radar. 
    \item SNR is \qty{30}{dB}.
    \item The subject has a velocity of \qty{0.05}{m/s} moving away from the radar.
    \item The AoA algorithm uses 512 bins to cover the $\pm 90\degree$ AoV, i.e. the angular resolution is $0.35\degree$.
\end{itemize}

The velocity of the subject is introduced following the assumption that a real person cannot stay absolutely stationary during the measurement. 
At a velocity of \qty{0.05}{m/s} and a frame time of \qty{50}{ms}, the total displacement will be \qty{2.5}{mm} and is considered negligible. 
The velocity provides a variation on the signal received at different chirps, as otherwise the multi-snapshot AoA estimation algorithms would receive an identical signal at all chirps and would yield a poor performance.

Combining the two DPCs with different AoA estimation algorithms, there are 14 methods in total to be evaluated. 
For each method, both the 1D search approach and the 2D sub-grid approach described in \Cref{sec:2d-bf} are included. 
For the 2D angle-FFT method, the full-grid approach is used instead of the sub-grid approach, since the benefit of the lower computational cost is less significant for FFTs.
The algorithms will be referred to using the format ``DPC-Method-1D/2D'' throughout the paper. 
For example, DPC1-Conv-2D refers to the conventional beamforming method in DPC1 that uses a 2D steering vector search. 
The angle-FFT method is not applicable in DPC2 as it is not a multi-snapshot algorithm. 
Algorithms in DPC1 include a CFAR peak detection step on the Range-Doppler heatmap, where the optimal parameters for the CFAR were searched on the training dataset. 
Then, the performance of the algorithms on the test dataset was evaluated and compared. 
The results are shown in \Cref{tab:c3-res-baseline} and \Cref{tab:c3-res-baseline-iou} as FMI and IoU (in \% and with the standard deviation in parentheses), respectively.

\begin{table}
\centering
\caption{FMI (standard deviation in parentheses) comparison between the algorithms when using a $4\times 4$ receiver array and a subject velocity of \qty{0.05}{m/s}.}
\label{tab:c3-res-baseline}
\begin{tabularx}{\linewidth}{|c| *{8}{Y|}}
\hline
\multirow{2}{*}{FMI in \%} &
  \multicolumn{2}{c|}{Angle-FFT} &
  \multicolumn{2}{c|}{Conv. BF} &
  \multicolumn{2}{c|}{MVDR BF} &
  \multicolumn{2}{c|}{MUSIC} \\ \cline{2-9} 
   & 1D & 2D & 1D & 2D & 1D & 2D & 1D & 2D \\ \hline
DPC1 & 68.3 \scriptsize{(7.5)}  & 68.2 \scriptsize{(7.9)}  & 60.6 \scriptsize{(8.7)}  & 67.2 \scriptsize{(7.6)}  & 67.7 \scriptsize{(7.6)}  & 74.5 \scriptsize{(6.7)}  & 69.7 \scriptsize{(6.9)}  & 77.0 
\scriptsize{(6.2)}  \\ \hline
DPC2 & \multicolumn{2}{c|}{NA}      & 43.7 \scriptsize{(7.8)}  & 46.5 \scriptsize{(7.1)}  & 50.2 \scriptsize{(7.6)} & 53.1 \scriptsize{(7.4)} & 52.7 \scriptsize{(7.4)} & 53.2 \scriptsize{(7.0)} \\ \hline  
\end{tabularx}
\end{table}

\begin{table}
\centering
\caption{IoU (standard deviation in parentheses) comparison between the algorithms when using a $4\times 4$ receiver array and a subject velocity of \qty{0.05}{m/s}.}
\label{tab:c3-res-baseline-iou}
\begin{tabularx}{\linewidth}{|c| *{8}{Y|}}
\hline
\multirow{2}{*}{IoU in \%} &
  \multicolumn{2}{c|}{Angle-FFT} &
  \multicolumn{2}{c|}{Conv. BF} &
  \multicolumn{2}{c|}{MVDR BF} &
  \multicolumn{2}{c|}{MUSIC} \\ \cline{2-9} 
   & 1D & 2D & 1D & 2D & 1D & 2D & 1D & 2D \\ \hline
DPC1 & 21.2 \scriptsize{(4.3)}  & 22.5 \scriptsize{(4.6)}  & 14.6 \scriptsize{(3.9)}  & 20.6 \scriptsize{(4.1)}  & 18.0 \scriptsize{(4.4)}  & 23.4 \scriptsize{(4.1)}  & 19.0 \scriptsize{(4.1)}  & 22.7 \scriptsize{(3.5)}  \\ \hline
DPC2 & \multicolumn{2}{c|}{NA}      & 11.2 \scriptsize{(3.2)}  & 12.2 \scriptsize{(3.0)}  & 13.2 \scriptsize{(3.3)} & 14.7 \scriptsize{(3.4)} & 14.6 \scriptsize{(3.2)} & 14.6 \scriptsize{(3.3)} \\ \hline
\end{tabularx}
\end{table}

\begin{figure}
    \centering
    \includegraphics[width=\linewidth]{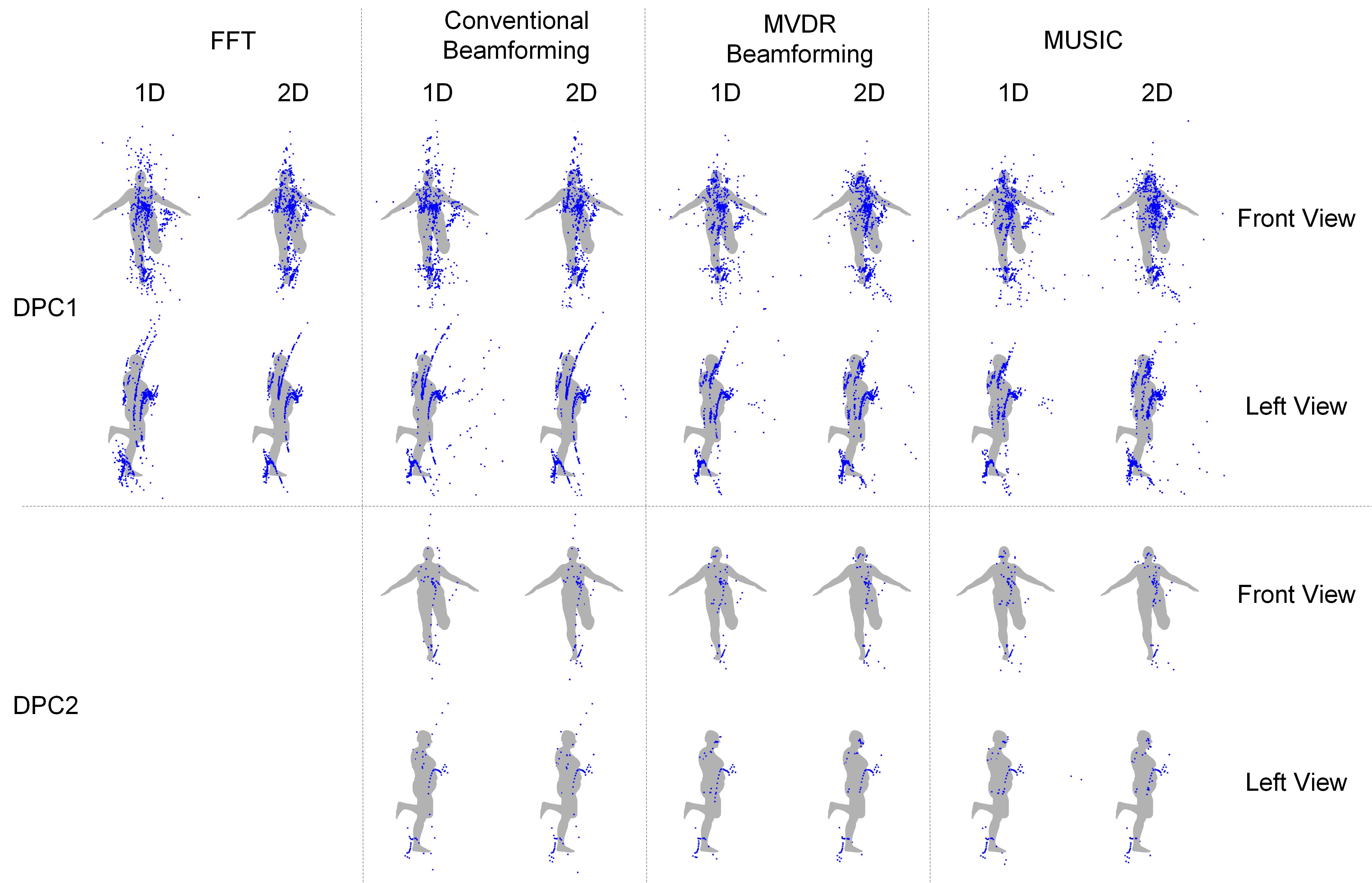}
    \caption{Examples of the radar detection using the different algorithms, when using a $4\times 4$ receiver array and a subject velocity of \qty{0.05}{m/s}.}
    \label{fig:c3-res-baseline-vis}
\end{figure}

There are a few important observations from the experiment. 
Even though the subject had a low velocity, the DPC1 with a Doppler-FFT outperformed the other significantly. 
One main reason is that, as the number of receivers is much lower than the number of signals, the AoA estimation algorithm can fail to distinguish points with a close angle.
Instead, these points will be identified as one strong signal source.
On the contrary, the CFAR peak detection step in DPC1 picks a set of points around the peak that are above the CFAR threshold.
As these points also contribute to the point cloud, the output becomes denser and the sensitivity is improved. 
This effect can be observed from the example detection shown in \Cref{fig:c3-res-baseline-vis}.

In terms of the different algorithms, the MVDR and MUSIC methods outperformed the angle-FFT and conventional methods, at the expense of higher complexity.
Meanwhile, all the 2D methods outperformed the 1D methods due to a more fine-grained resolution (as shown earlier in \Cref{fig:2d-bf}).
The best performance was achieved with the DPC1-MVDR-2D and DPC1-MUSIC-2D methods, with an FMI of 74.5\% and 77.0\%, respectively.
However, the IoU metrics show that the point clouds were still far from the objective of high-accuracy scene reconstruction, as the highest IoU was only 23.4\%. 
It can be seen from \Cref{fig:c3-res-baseline-vis} that, while the distribution of the point cloud mostly fitted the subject, the distribution was not even and there were body parts (like the hands) that received fewer points. 
Therefore, there is still a big gap before the radar output can be directly used by applications that require high quality data.

\Cref{tab:c3-res-baseline-time} compares the algorithms in terms of execution time. 
The algorithms were run using the same dataset and parameters multiple times. 
The algorithms were written in Python without any processor-specific optimization and were run on one Intel i7-9700K CPU core. 
The result is shown as the relative execution time of each algorithm when compared with the DPC1-FFT-1D method (the most lightweight method) and normalized with the number of detected points, to give an indication of their relative complexity.
All the 2D methods have a higher complexity than the 1D methods.
For algorithms in DPC1, the 1D angle-FFT method has the lowest computational cost. 
With the sub-grid optimization, the complexity of the 2D beamforming and MUSIC methods can be kept at around twice the 1D methods.
The complexity without the sub-grid optimization is expected to be much higher, as can be estimated from the difference between the 2D and 1D angle-FFT methods. 
When considering both the complexity and the performance, the DPC1-FFT-1D method provides a good trade-off between them.
The MVDR methods and MUSIC methods in DPC1 give the best performance at the cost of 9x execution time and require additional efforts on the hardware and implementation.
It is worth noting that many mmWave radar systems, like \cite{my-posture,mmpose,mmwave-radhar}, are built based on the DPC1-FFT-1D method. 
Therefore, these systems can potentially benefit from a more complex AoA estimation algorithm.

\begin{table}
\centering
\caption{Normalized execution time comparison between the algorithms using the baseline setup.}
\label{tab:c3-res-baseline-time}
\begin{tabularx}{\linewidth}{|c| *{8}{Y|}}
\hline
\multirow{2}{*}{\begin{tabular}[c]{@{}c@{}}Normalized\\Complexity\end{tabular}} &
  \multicolumn{2}{c|}{Angle-FFT} &
  \multicolumn{2}{c|}{Conv. BF} &
  \multicolumn{2}{c|}{MVDR BF} &
  \multicolumn{2}{c|}{MUSIC}  \\ \cline{2-9} 
   & 1D & 2D & 1D & 2D & 1D & 2D & 1D & 2D \\ \hline
DPC1 & 1.00  & 13.42  & 4.38  & 9.32  & 3.51  & 8.99  & 4.02  & 8.85  \\ \hline
DPC2 & \multicolumn{2}{c|}{NA}      & 5.69  & 12.38  & 5.31 & 10.89 & 5.67 & 10.61 \\ \hline
\end{tabularx}
\end{table}

\subsection{Subject Velocity}\label{sec:pc-velocity}
The motion of the subject being sensed has a significant impact on the detection output. 
In DPC1, a higher velocity makes a subject easier to be identified in the Range-Doppler heatmap. 
Due to the relative position difference between the body parts of the subject, they will have a different radial velocity with respect to the radar, making them distinguishable in the Range-Doppler heatmap. 
In DPC2, a higher velocity increases the variance of the signal between chirps and allows a better estimate of the data covariance matrix. 
To verify the theorem, an experiment was carried out using the same configuration as the baseline setup, except that the velocity of the subject was set to different values from \qtyrange{0.1}{1}{m/s}. 
The ground truth point cloud was taken as the average position of the subject during the motion.

\begin{figure}
    \centering
    \includegraphics[width=0.5\linewidth]{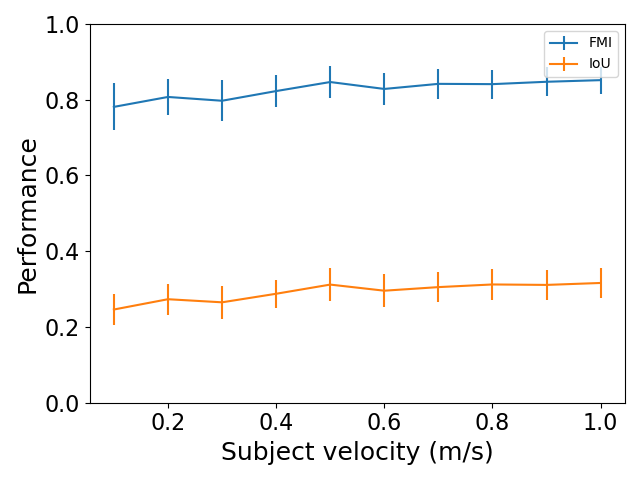}
    \caption{FMI and IoU (with errors) of the DPC1-MUSIC-2D algorithm with different subject velocities.}
    \label{fig:fmi-velocity}
\end{figure}

\begin{figure}
    \centering
    \includegraphics[width=\linewidth]{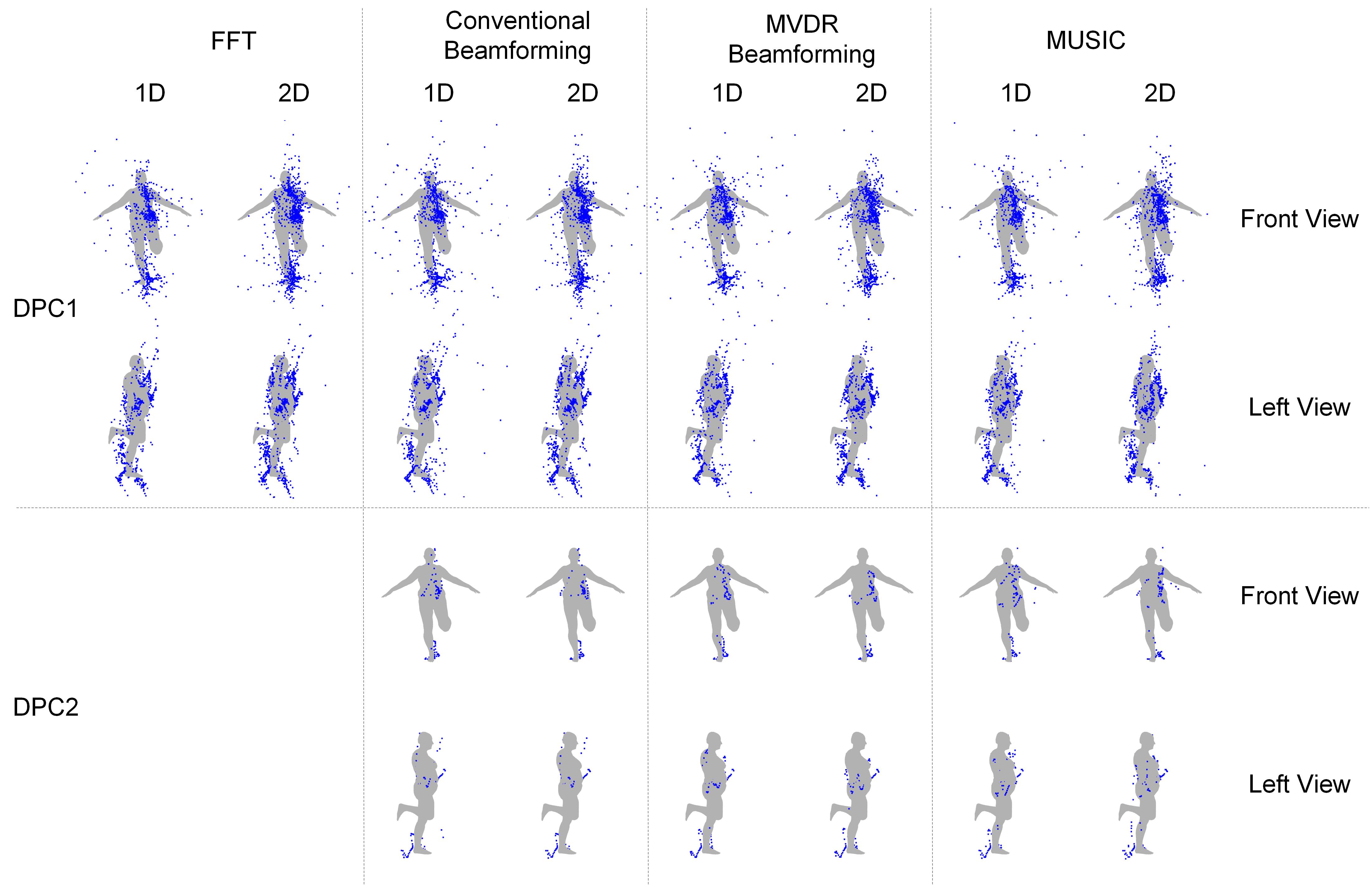}
    \caption{Examples of the radar detection using the different algorithms, when using a $4\times 4$ receiver array and a subject velocity of \qty{1}{m/s}.}
    \label{fig:c3-res-v1-vis}
\end{figure}

\Cref{tab:c3-res-v0.5} and \Cref{tab:c3-res-v1} show two examples of the experiment where the subject velocity was set to \qty{0.5}{m/s} and \qty{1}{m/s}, respectively. 
When compared with \Cref{tab:c3-res-baseline}, all algorithms achieved a 2.6\% to 14.5\% improvement in terms of the FMI when the subject had an increased velocity. 
\Cref{fig:fmi-velocity} shows the FMI and IoU of the DPC1-MUSIC-2D method with different subject velocities from \qtyrange{0.1}{1}{m/s}.
An overall positive correlation can be observed between the subject velocity and the detection performance, and the impact is the most obvious at lower velocities (around \qty{0.5}{m/s}). 
Some examples of the detection at \qty{1}{m/s} are shown in \Cref{fig:c3-res-v1-vis}.

\begin{table}
\centering
\caption{Relative FMI difference of the algorithms when using a $4\times 4$ receiver array and a subject velocity of \qty{0.5}{m/s} in comparison to \qty{0.05}{m/s}.}
\label{tab:c3-res-v0.5}
\begin{tabularx}{\linewidth}{|c| *{8}{Y|}}
\hline
\multirow{2}{*}{FMI in \%} &
  \multicolumn{2}{c|}{Angle-FFT} &
  \multicolumn{2}{c|}{Conv. BF} &
  \multicolumn{2}{c|}{MVDR BF} &
  \multicolumn{2}{c|}{MUSIC}  \\ \cline{2-9} 
   & 1D & 2D & 1D & 2D & 1D & 2D & 1D & 2D \\ \hline
DPC1 & +8.3   & +11.4   & +12.1   & +10.7   & +10.4   & +8.8   & +10.1   & +7.7   \\ \hline 
DPC2 & \multicolumn{2}{c|}{NA}      & +4.6   & +2.6   & +5.3  & +4.6  & +8.4  & +5.6  \\ \hline
\end{tabularx}
\end{table}

\begin{table}
\centering
\caption{Relative FMI difference of the algorithms when using a $4\times 4$ receiver array and a subject velocity of \qty{1}{m/s} in comparison to  \qty{0.05}{m/s}.}
\label{tab:c3-res-v1}
\begin{tabularx}{\linewidth}{|c| *{8}{Y|}}
\hline
\multirow{2}{*}{FMI in \%} &
  \multicolumn{2}{c|}{Angle-FFT} &
  \multicolumn{2}{c|}{Conv. BF} &
  \multicolumn{2}{c|}{MVDR BF} &
  \multicolumn{2}{c|}{MUSIC} \\ \cline{2-9} 
   & 1D & 2D & 1D & 2D & 1D & 2D & 1D & 2D \\ \hline
DPC1 & +9.7   & +13.0   & +14.5   & +12.8   & +12.0   & +9.8   & +11.6   & +8.2   \\ \hline 
DPC2 & \multicolumn{2}{c|}{NA}      & +5.3   & +2.6   & +4.3  & +3.4  & +9.3  & +5.8  \\ \hline
\end{tabularx}
\end{table}

\subsection{SNR}\label{sec:pc-noise}
In a practical environment, a radar system can experience noise from different sources, such as the thermal noise of the radar chip.
The SNR also depends on the distance between the radar and the subject, as the signal power drops quickly along with the distance. 
In the simulator, the SNR can be controlled by the power of the noise term $n$ in \Cref{eq:if-simulation}.
In this section, the performance of the algorithms between a high SNR environment (\qty{40}{dB}) and a lower SNR environment (\qty{5}{dB}) is compared. 
Two experiments were carried with the subject velocity set to \qty{0.05}{m/s} and \qty{0.5}{m/s}, respectively.
The results are shown in \Cref{tab:c3-res-lowsnr} and \Cref{tab:c3-res-lowsnr-v0.5}.

\begin{table}
\centering
\caption{Performance difference when using a $4\times 4$ receiver array and a subject velocity of \qty{0.05}{m/s} in a low SNR environment (\qty{5}{dB} in comparison to \qty{30}{dB}).}
\label{tab:c3-res-lowsnr}
\begin{tabularx}{\linewidth}{|c| *{8}{Y|}}
\hline
\multirow{2}{*}{FMI in \%} &
  \multicolumn{2}{c|}{Angle-FFT} &
  \multicolumn{2}{c|}{Conv. BF} &
  \multicolumn{2}{c|}{MVDR BF} &
  \multicolumn{2}{c|}{MUSIC}  \\ \cline{2-9} 
   & 1D & 2D & 1D & 2D & 1D & 2D & 1D & 2D \\ \hline
DPC1 & -8.1   & -5.8   & -5.5   & -5.6   & -6.4   & -6.3   & -5.7   & -6.2   \\ \hline 
DPC2 & \multicolumn{2}{c|}{NA}      & +2.2   & +2.8   & +1.7  & +2.7  & +1.6  & +2.4  \\ \hline
\end{tabularx}
\end{table}

\begin{table}
\centering
\caption{Performance difference when using a $4\times 4$ receiver array and a subject velocity of \qty{0.5}{m/s} in a low SNR environment (\qty{5}{dB} in comparison to \qty{30}{dB}).}
\label{tab:c3-res-lowsnr-v0.5}
\begin{tabularx}{\linewidth}{|c| *{8}{Y|}}
\hline
\multirow{2}{*}{FMI in \%} &
  \multicolumn{2}{c|}{Angle-FFT} &
  \multicolumn{2}{c|}{Conv. BF} &
  \multicolumn{2}{c|}{MVDR BF} &
  \multicolumn{2}{c|}{MUSIC}  \\ \cline{2-9} 
   & 1D & 2D & 1D & 2D & 1D & 2D & 1D & 2D \\ \hline
DPC1 & -7.8   & -6.2   & -7.2   & -6.1   & -8.3   & -7.5   & -8.9   & -7.4   \\ \hline 
DPC2 & \multicolumn{2}{c|}{NA}      & +2.5   & +2.8   & +1.0  & +0.8  & -0.7  & +0.9  \\ \hline
\end{tabularx}
\end{table}

In the low SNR environment, all the algorithms in DPC1 experienced a similar drop in performance, as expected.
However, the algorithms in the DPC2 showed a higher performance. 
The reason is that the higher noise affected the model order estimation step and the system tends to report a higher number of points. 
Taking the DPC2-Conv-2D method as an example, the average size of the detected point cloud was found to be 20.3\% higher in a low SNR environment than in a higher SNR environment. 
However, this was still insufficient to reach a similar performance as DPC1.


\subsection{Antenna Layout}\label{sec:pc-antenna}
Theoretically, the antenna layout determines the angular resolution that an AoA estimation algorithm can achieve. 
The more receivers in one direction, the higher resolution the radar can measure \cite{mimo}. 
However, this is questionable when the signal sources are spatially close and continuous.
Meanwhile, having more antennas also increases the cost of the hardware, as more circuit components, processing units and memory would be required.
Therefore, it is beneficial to study the relationship between the antenna layout and the output quality and find the optimal trade-off for an application.

\begin{figure}
    \centering
    \includegraphics[width=0.75\linewidth]{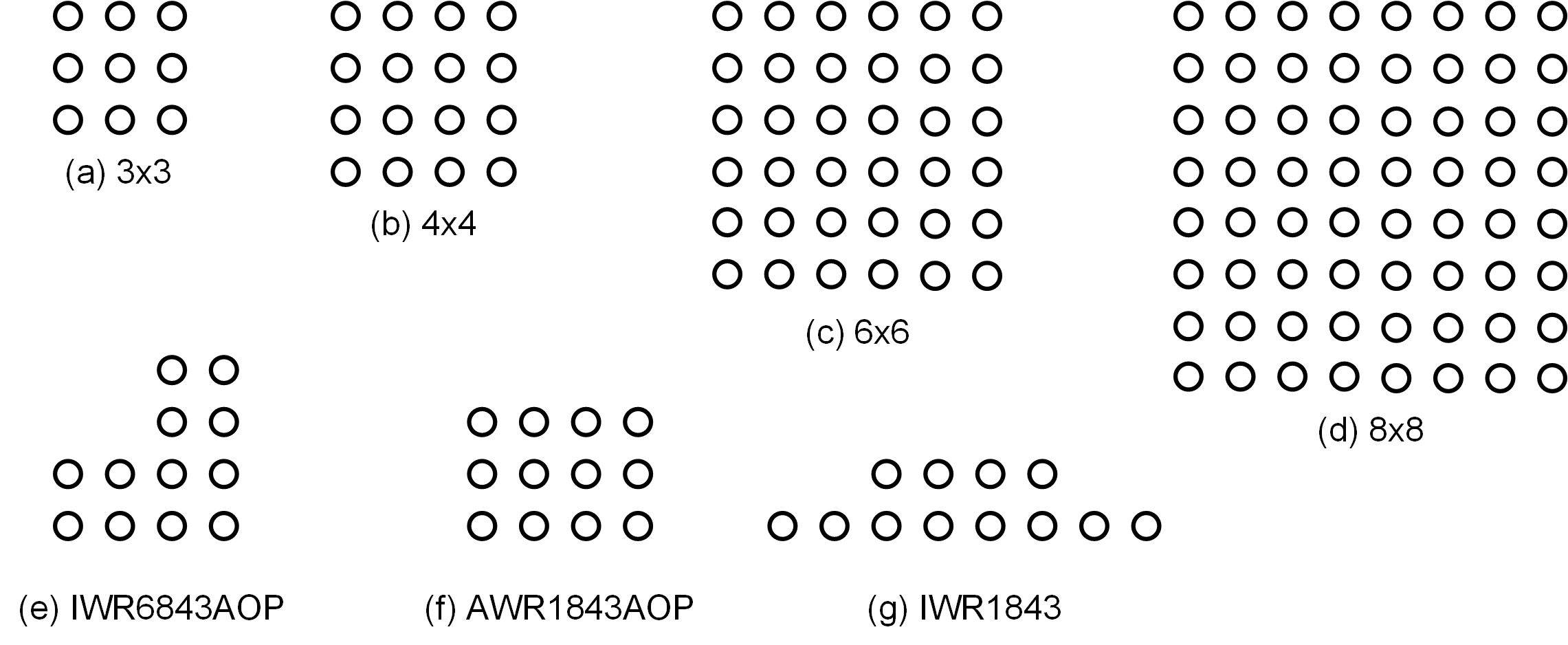}
    \caption{The list of receiver layouts being evaluated. (a)-(d) are square antenna arrays. (e)-(f) are non-regular antenna arrays implemented on TI radars. }
    \label{fig:pc-antenna-layout}
\end{figure}

Common commercial mmWave radars use up to three transmitters and up to four receivers, giving up to twelve virtual receivers as a receiving array.
Some radar models are designed for automotive applications and prioritize the azimuth direction, while others are designed for general purpose applications and have a similar resolution in both the azimuth and elevation directions. 
In this section, common antenna layouts implemented on the TI radars are evaluated and compared, as well as a few square-shape antenna layouts that are more common in research projects, as listed in \Cref{fig:pc-antenna-layout}. 
The same radar configuration and scene setup in \Cref{sec:pc-baseline} were used. 
The experiment compares the antenna layouts using the DPC1-MUSIC-2D algorithm (the best performing algorithm). 
The result is shown in \Cref{tab:c3-res-antenna}.

\begin{table}
\centering
\caption{Performance comparison between different antenna layouts using the baseline configuration and the DPC1-MUSIC-2D algorithm (standard deviation in parentheses).}
\label{tab:c3-res-antenna}
\begin{tabular}{|c|c|c|c|c|c|c|c|}
\hline
\begin{tabular}[c]{@{}c@{}}Antenna\\ Layouts\end{tabular} &
  a &
  b &
  c &
  d &
  e &
  f &
  g \\ \hline
\begin{tabular}[c]{@{}c@{}}FMI\\ in \%\end{tabular} &
  \begin{tabular}[c]{@{}c@{}}76.7\\ \scriptsize{(6.4)}\end{tabular} &
  \begin{tabular}[c]{@{}c@{}}77.0\\ \scriptsize{(6.2)}\end{tabular} &
  \begin{tabular}[c]{@{}c@{}}76.8\\ \scriptsize{(4.8)}\end{tabular} &
  \begin{tabular}[c]{@{}c@{}}77.9\\ \scriptsize{(4.5)}\end{tabular} &
  \begin{tabular}[c]{@{}c@{}}72.4\\ \scriptsize{(6.3)}\end{tabular} &
  \begin{tabular}[c]{@{}c@{}}77.8\\ \scriptsize{(5.9)}\end{tabular} &
  \begin{tabular}[c]{@{}c@{}}65.0\\ \scriptsize{(6.0)}\end{tabular} \\ \hline
\begin{tabular}[c]{@{}c@{}}IoU\\ in \%\end{tabular} &
  \begin{tabular}[c]{@{}c@{}}23.4\\ \scriptsize{(4.1)}\end{tabular} &
  \begin{tabular}[c]{@{}c@{}}22.7\\ \scriptsize{(3.5)}\end{tabular} &
  \begin{tabular}[c]{@{}c@{}}20.5\\ \scriptsize{(2.8)}\end{tabular} &
  \begin{tabular}[c]{@{}c@{}}18.8\\ \scriptsize{(2.7)}\end{tabular} &
  \begin{tabular}[c]{@{}c@{}}20.8\\ \scriptsize{(4.0)}\end{tabular} &
  \begin{tabular}[c]{@{}c@{}}23.9\\ \scriptsize{(4.2)}\end{tabular} &
  \begin{tabular}[c]{@{}c@{}}17.0\\ \scriptsize{(3.5)}\end{tabular} \\ \hline
\end{tabular}
\end{table}

\begin{figure}
    \centering
    \includegraphics[width=\linewidth]{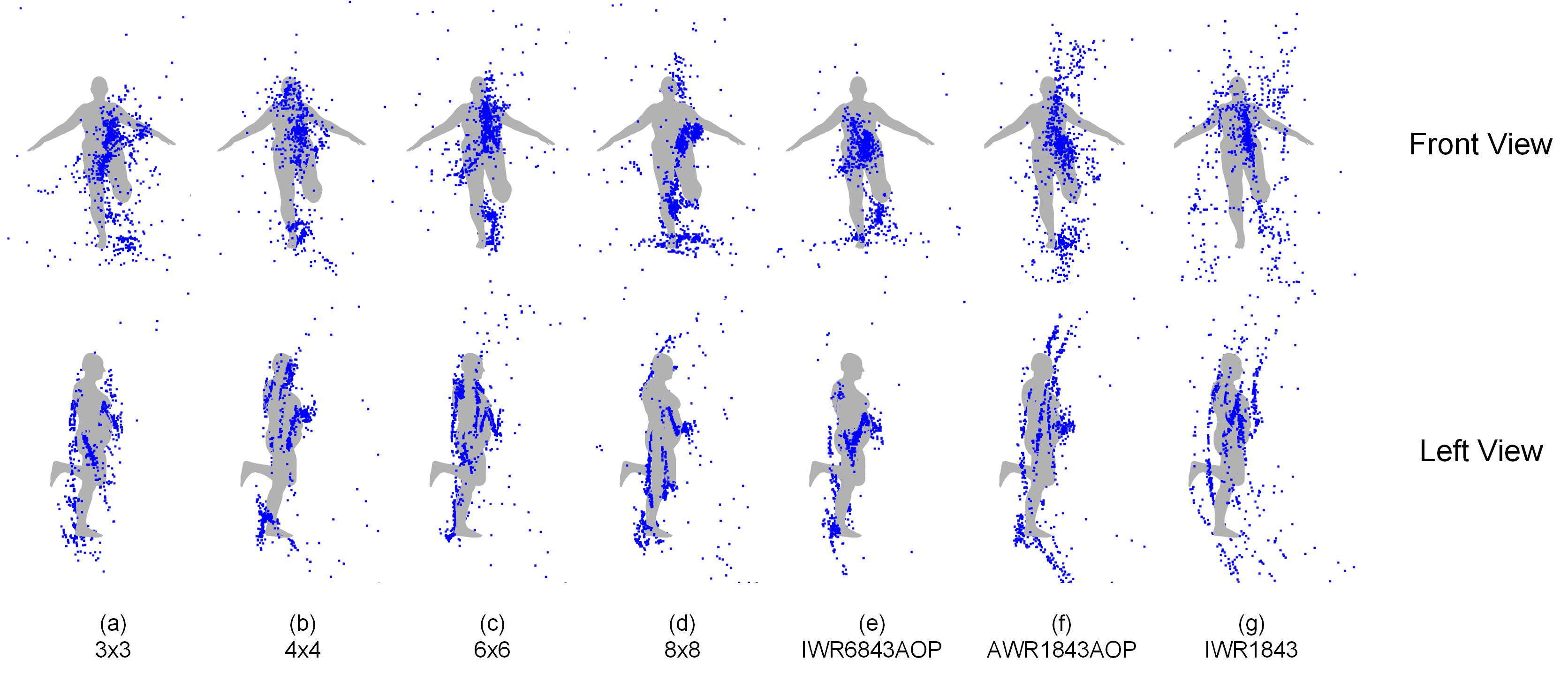}
    \caption{Examples of the radar detection using the different antenna layouts with the baseline setup.}
    \label{fig:c3-res-antenna-vis}
\end{figure}

It can be seen that most antenna layouts had similar performance, except the layout (g) which had a worse performance as it is designed for automotive applications.
The layout (e) has a non-uniform antenna distribution that slightly affected its performance. 
All other layouts showed a similar performance regardless of the antenna size. 
Therefore, considering the increased hardware cost and computational cost of increasing the number of antennas, a small antenna size can be preferable for 3D sensing applications. 
\Cref{fig:c3-res-antenna-vis} shows some examples of the detection using different antenna layouts. 

\subsection{Chirp Configuration}
The chirp configuration can have various effects on the distance detection and velocity detection.
These factors can indirectly affect the quality of the final point cloud. 
In this section, three different chirp configurations were tested and compared against the baseline configuration in \Cref{sec:pc-baseline}.
The details of the three configurations (named A, B and C) and the performance are shown in \Cref{tab:c3-res-chirp}.
Each configuration has certain parameter cut to 80\% to evaluate the effect on the output. 
Configuration A had an 80\% reduced chirp slope and, hence, a reduced effective bandwidth from \qty{4}{GHz} to \qty{3.2}{GHz}.
Configuration B had an 80\% reduced ADC sampling rate that reduced the samples per chirp from 1500 to 1200.
Configuration C had an 80\% reduced number of chirps per frame, from 50 to 40. 
All other parameters were kept the same as the baseline with the DPC1-MUSIC-2D algorithm. 

\begin{table}
\centering
\caption{FMI (standard deviation in parentheses) comparison between four chirp configurations using the DPC1-MUSIC-2D algorithm.}
\label{tab:c3-res-chirp}
\begin{tabularx}{\linewidth}{|c| *{4}{Y|}}
\hline
Chirp Configuration                                                   & Baseline                                                                           & A                                                                                  & B                                                                                  & C                                                                                  \\ \hline
Slope of the chirp (MHz/us) & 40                                                & 32                                                & 40                                                & 40                                                \\ \hline
ADC sampling rate (MHz)      & 15                                                   & 15                                                   & 12                                                   & 15                                                   \\ \hline
Chirps per frame                                                      & 50                                                                                 & 50                                                                                 & 50                                                                                 & 40                                                                                 \\ \hline
FMI in \%                                                             & \begin{tabular}[c]{@{}c@{}}77.0\\ \scriptsize{(6.2)}\end{tabular} & \begin{tabular}[c]{@{}c@{}}71.1\\ \scriptsize{(6.8)}\end{tabular} & \begin{tabular}[c]{@{}c@{}}76.5\\ \scriptsize{(6.0)}\end{tabular} & \begin{tabular}[c]{@{}c@{}}70.2\\ \scriptsize{(6.0)}\end{tabular} \\ \hline
\end{tabularx}
\end{table}

The result shows that the performance can be strongly affected by the effective bandwidth and the number of chirps. 
The former affects the distance resolution of the detection, and the latter affects the Doppler resolution.
Reducing either of these parameters reduces the accuracy of the range-Doppler heatmap and the estimation of the covariance matrix. 
On the other hand, the effect of reducing the ADC sampling rate and the number of samples per chirp is much less significant. 

\section{Super-resolution Point Cloud Construction Algorithm}\label{sec:srpc}
It can be seen from \Cref{fig:c3-res-baseline-vis} and \Cref{fig:c3-res-v1-vis} that the constructed point clouds can be noisy and the distribution of the points can be imbalanced.
One major reason is that the point cloud construction relies on the peak detection result over the range-Doppler-FFT spectrum, so the distribution of the points will be limited by the resolution of the FFT, and the points will have a discrete distribution in the range domain (as the curve-like data from the left view).
Although it is possible to improve this resolution, such as zero padding the data before applying the FFT, it would also increase the computational cost and memory consumption. 
Meanwhile, there are false detected points due to the outliers from the peak detection stage. To address the mentioned issue and improve the quality of the constructed point cloud, a novel super-resolution point cloud construction (SRPC) algorithm is proposed.

\begin{figure}
    \centering
    \begin{subfigure}[t]{0.49\linewidth}
        \includegraphics[width=\linewidth]{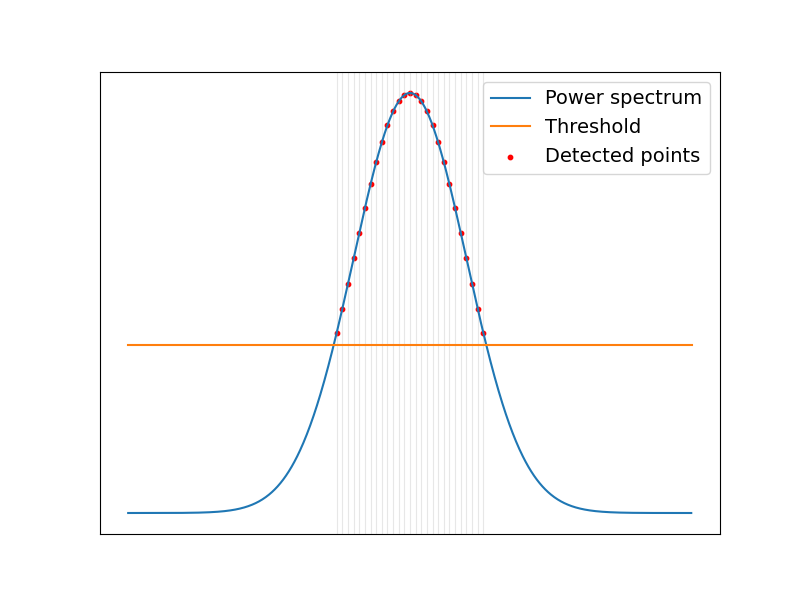}
        \caption{Points detected without SRPC.}
    \end{subfigure}
    \begin{subfigure}[t]{0.49\linewidth}
        \includegraphics[width=\linewidth]{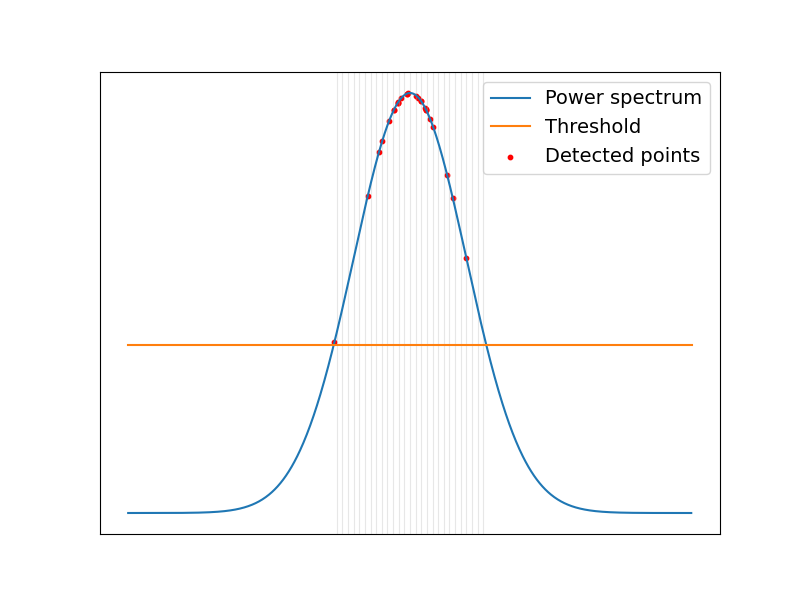}
        \caption{Points detected with SRPC.}
    \end{subfigure}
    \caption{Using SRPC algorithm to improve the resolution and distribution of the data.}
    \label{fig:srpc}
\end{figure}

The SRPC algorithm aims to improve the distribution of the point cloud and make it span more naturally in the spatial space. 
The rationale is shown in \Cref{fig:srpc}.
When detecting peaks in a range-Doppler spectrum or an angle spectrum, a common approach is taking all points above a static or dynamic threshold, where the distribution of the points is limited by the resolution of the original data.
An example of this effect is shown in \Cref{fig:srpc}a, where the grid represents the resolution of the data and all the detected points must fall on the grid.
The SRPC algorithm aims to return a set of points that have a higher resolution than the original data and fall more naturally on the distribution curve, as shown in \Cref{fig:srpc}b.

The algorithm can be broken down into the following steps. 
First, the power spectrum is upsampled into the desired resolution using linear interpolation.
Then, for each of the originally detected points $i\in [1..K]$, the algorithm randomly samples $n_i$ points around it with a probability distribution being the amplitude of the upsampled power spectrum.
The value of $n_i$ is calculated as:
\begin{equation}
    n_i=\frac{p_i\cdot \alpha_{SRPC}}{th}
\end{equation}
where $p_i$ is the power of the point, $th$ is the threshold of the peak detection algorithm, and $\alpha_{SRPC}$ is a global hyperparameter that controls the aggressiveness of the algorithm.
The term $p_i$ ensures that a point with higher power will be sampled into more points, as the power indicates the confidence that a point can represent a real signal source.
The parameter $\alpha_{SRPC}$ amplifies the importance of $p_i$, where a higher $\alpha_{SRPC}$ pushes the distribution of the points towards the peak of the spectrum and gives a more dense distribution.
The sampling process is repeated for each point $i$ to form a new point list.
Finally, $K$ points (the population of the original detection) are randomly selected from the new point list, so that the total number of detected points is kept the same and the computational cost of the rest of the system is not affected. 
Since the algorithm tends to sample more points at higher power, the distribution of the final points will also tend to be around higher powers, and, hence, gives a more natural distribution regarding the power spectrum and overcomes the limitation of the original data resolution. 
The time complexity of the SRPC algorithm is approximately $O(K\cdot n_i)$, where a typical value of $n_i$ can fall between 2 and 8.

When constructing the point cloud, the SRPC is applied when detecting peaks from the range-FFT spectrum and detecting peaks from the angle spectrum in the AoA estimation step.
The former improves the data distribution in the range domain and eliminates the curve-like effect when looking at the point cloud from the left view. 
The latter improves the data distribution in the angle domain so that the points tend to span into the space rather than appearing as a dense cluster.
Meanwhile, since the points will be distributed around higher powers, the probability of outliers will be reduced. 

\begin{figure}
    \centering
    \includegraphics[width=\linewidth]{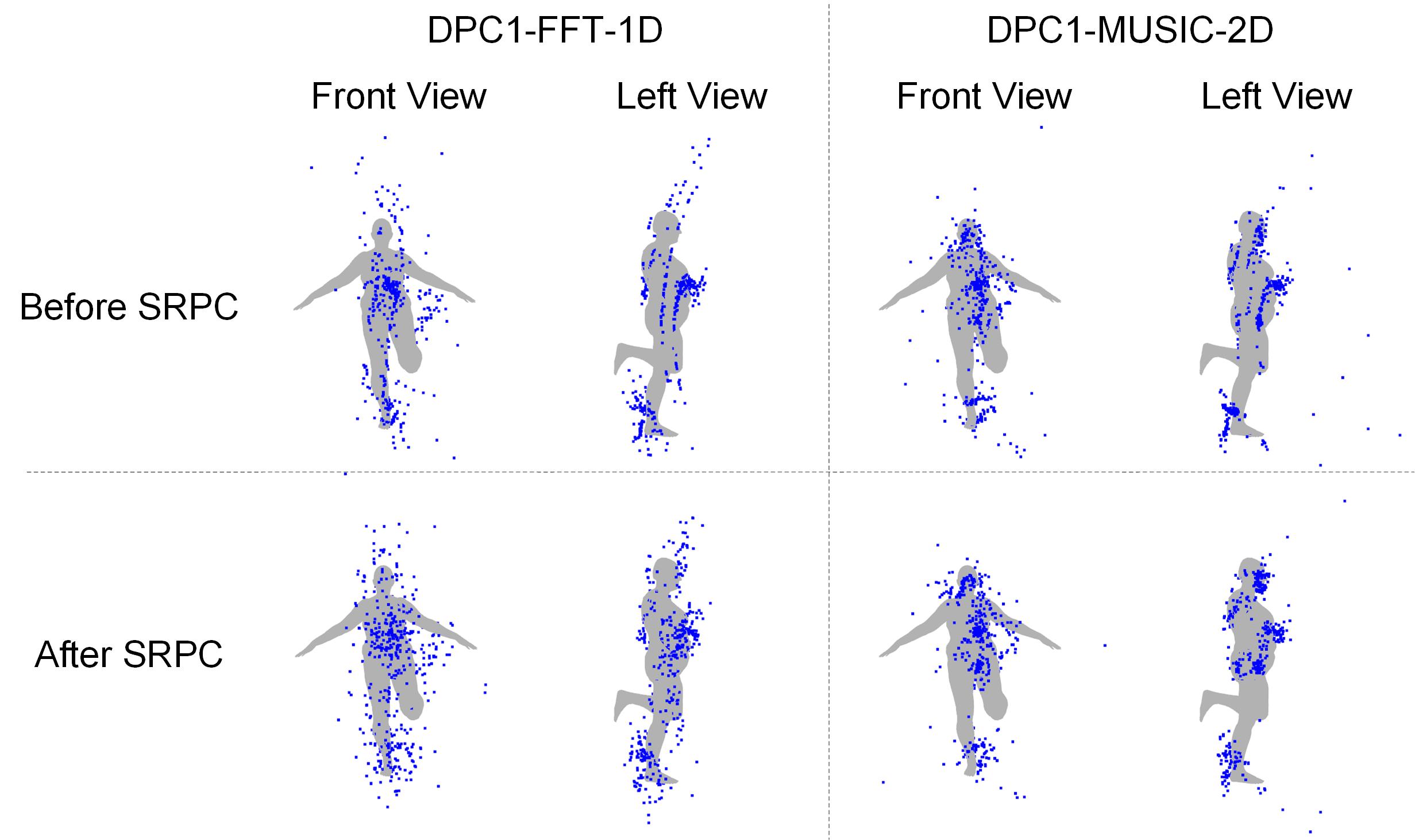}
    \caption{Examples of point clouds constructed with and without the SRPC algorithm.}
    \label{fig:srpc-vis}
\end{figure}

To evaluate the proposed SRPC algorithm, it was inserted into the DPC1-FFT-1D and DPC1-MUSIC-2D methods mentioned in \Cref{sec:pc-baseline} when using the baseline setup. 
The two methods were chosen as they represent the most lightweight algorithm and the most accurate algorithm, respectively. 
Since the SRPC is likely to produce point clouds with different sizes and to ensure a fair comparison, a fixed number of 512 points were randomly taken from the point cloud generated by each algorithm for the evaluation. 
The result is shown in \Cref{fig:srpc-vis}.
After applying the SRPC algorithm, the distribution of the point cloud appeared to be more natural and better distributed around the ground truth, and the outliers in the original detection were reduced. 
A quantitative evaluation is shown in \Cref{tab:srpc-res}.
The performance without SRPC dropped slightly when compared with \Cref{tab:c3-res-baseline} because the output size was forced to be 512, but both metrics have improved after applying SRPC.
Therefore, it is shown that the SRPC algorithm can successfully improve the data point distribution, reduce the outliers and produce a more natural point cloud that can be potentially preferable for higher-level applications.
Future work of this research includes an efficient hardware implementation of this algorithm using the radar on-chip processors so that it can be further verified in real-world scenarios, as well as an evaluation of its effectiveness in higher-level applications like posture estimation. 

\begin{table}
\centering
\caption{Performance comparison of two algorithms with and without SRPC.}
\label{tab:srpc-res}
\begin{tabularx}{0.8\linewidth}{|c| *{4}{Y|}}
\hline
\multirow{2}{*}{} & \multicolumn{2}{c|}{DPC1-FFT-1D}  & \multicolumn{2}{c|}{DPC1-MUSIC-2D}     \\ \cline{2-5}
                  & FMI         & IoU  & FMI           & IoU  \\ \hline
Without SRPC      & 64.9        & 20.2 & 72.1          & 22.9 \\ \hline
With SRPC         & 69.5        & 23.6 & 72.9          & 25.9 \\ \hline
\end{tabularx}
\end{table}

\section{Conclusion}\label{sec:conclusion}
In this paper, a mmWave radar simulator is presented. 
The system is used to evaluate the ability of the mmWave radar as a 3D imaging sensor. 
A mmWave radar dataset is constructed using the FAUST dataset as the ground truth to provide 3D mesh models of human subjects, from which mmWave radar IF signals are simulated and used to evaluate different point cloud construction algorithms. 
The FMI and IoU metrics are defined to evaluate the quality of the generated point cloud. 
The evaluation is performed regarding a set of different factors, including the DPCs, AoA estimation algorithms, subject velocity, SNR, antenna layout and chirp configuration. 
It was found that the DPC combining a range-Doppler-FFT and a single-snapshot AoA estimation algorithm gives better performance. 
Among all the AoA estimation algorithms, the angle-FFT method gives a good trade-off between high performance and low computational cost, whereas the more advanced AoA estimation algorithms, like MVDR and MUSIC, give the best performance at up to 9x higher execution time. 
The velocity of the subject helps significantly in the detection, as the algorithms are better at detecting a moving subject than a stationary object.
When comparing common antenna layouts, large square antenna arrays give the best performance, but the advantage is not significant in a 3D imaging application when the data sources are spatially close and continuous. 
It is shown that the performance of the point cloud detection benefits from higher effective bandwidth and a higher number of chirps per frame. 
Finally, a novel SRPC algorithm is proposed for improving the resolution and distribution of the point cloud and reducing the probability of outliers. 
The algorithm applies to the range-Doppler-FFT peak detection stage and the AoA estimation stage and detects points at a higher resolution that fits the power spectrum better.
When evaluating the algorithm using the simulation system, it has been shown that the algorithm can successfully improve the data distribution and produces a more natural point cloud. 

\balance
\bibliographystyle{IEEEtran}
\bibliography{ref}

\clearpage
\nobalance

\appendices
\crefalias{section}{appendix}
\section{IF Signal}\label{apx:if}

In a typical FMCW radar model, the transmitter sends a chirp signal $S_{tx}$ (a signal with frequency increasing linearly with time) to detect any object in front of the radar. 
When $S_{tx}$ is reflected by the object, the signal is received as $S_{rx}$.
Assuming the signal has an initial frequency $f_0$ and a slope of $S$, then the frequency of $S_{tx}$ is a function of $t$:
\begin{equation}
f_{tx}(t) = f_0 + S\cdot t
\end{equation}
The instantaneous phase of the signal is a function of $t$ and is the integral of $f_{tx}$:
\begin{equation}\label{eq:tx}
\begin{split}
\phi_{tx}(t)  &= \int_{\tau=0}^t 2\pi \cdot f_{tx}(\tau) \ d\tau \\
        &= \int_{\tau=0}^t 2\pi \cdot (f_0 + S\cdot \tau) \ d\tau \\
        &= 2\pi\cdot f_0\cdot t + \int_{\tau=0}^t 2\pi\cdot S\cdot \tau \ d\tau \\
        &= 2\pi\cdot f_0\cdot t + 2\pi\cdot \frac{1}{2} S\cdot t^2 \\
        &= 2\pi\cdot f_0\cdot t + \pi\cdot S\cdot t^2
\end{split}
\end{equation}
The transmitted signal $S_{tx}$ can be written as a sinusoid signal:
\begin{equation}\label{eq:rx}
S_{tx}(t) = A\cdot cos(2\pi f_0 t + \pi S t^2)
\end{equation}
where $A$ is the transmission power. 
The received signal is a delayed and downscaled version of $S_{tx}$:
\begin{equation}
S_{rx}(t) = \alpha A\cdot cos\big(2\pi f_0 (t-\tau) + \pi S (t-\tau)^2\big)
\end{equation}
where $\tau$ is the ToF of the signal and indicates the distance of the object, and $\alpha$ is the downscale factor that models the transmission loss.
The two signals, $S_{tx}$ and $S_{rx}$, are combined through a mixer (a multiplier) to generate one signal with both the sum frequency and the difference frequency:
\begin{equation}\label{eq:mixer}
\begin{split}
S_{tx}(t)\cdot& S_{rx}(t) \\
= \frac{\alpha A^2}{2}  \Big(&
        cos\big(
        (2\pi f_0 t + \pi S t^2)  +  (2\pi f_0 (t-\tau) + \pi S (t-\tau)^2)
        \big) + \\&
        cos\big(
        (2\pi f_0 t + \pi S t^2)  -  (2\pi f_0 (t-\tau) + \pi S (t-\tau)^2)
        \big) 
        \Big)\\
        = \frac{\alpha A^2}{2}  \Big(&
        cos\big(
        2\pi (2f_0-S\tau) t + 2\pi S t^2 + \pi S \tau^2  -2\pi f_0 \tau
        \big) + \\&
        cos\big(
        2\pi (S\tau) t +2\pi f_0 \tau - \pi S \tau^2  
        \big) 
        \Big)        
\end{split}
\end{equation}
There are two $cos$ terms in the result. 
The first one has a frequency of $2f_0$ and will be removed by a low pass filter.
The second one is called the IF signal or the beat frequency. 
The IF signal has the equation:
\begin{equation}
    IF(t) = B \cdot cos\big(
        2\pi (S\tau) t +2\pi f_0 \tau - \pi S \tau^2  
        \big) 
\end{equation}
where $B=\frac{\alpha A^2}{2}$.
The signal has a frequency $S\tau$, i.e. the slope of the chirp multiplied by the ToF.
Therefore, the frequency of the IF signal is directly proportional to the ToF.
Given that the slope of the chirp $S$ is known, the distance of the object can be calculated from the frequency of the IF signal. 
The phase of the IF signal, $(2\pi f_0 \tau - \pi S \tau^2)$, can be simplified to $(2\pi f_0 \tau)$, as the second term is negligible: $S$ has an order of $10^{12}$, $\tau$ has an order of $10^{-8}$, so $(\pi S \tau^2)$ will have a negligible order of $10^{-4}$.
In summary, the IF signal can be written as:
\begin{equation}\label{eq:IF1}
    IF(t) = B \cdot cos(\omega_b t + \phi_b)
\end{equation}
where the angular frequency $\omega_b$ and the phase of the signal $\phi_b$ are:
\begin{equation}\label{eq:IF2}
    \omega_b = 2\pi\cdot S\tau,  \quad \phi_b = 2\pi f_0\tau
\end{equation}

The above equations assume that the object is stationary.
If the object is moving, the ToF $\tau$ will be varying with respect to $t$. 
However, considering that this variation is limited to a single chirp time, it is unlikely to produce a big change in the frequency. 
The change in phase can be more significant, but will only affect certain applications where the phase information is critical, such as vital sign monitoring. 
In such cases, the phase can be written as a function of $t$ as $\phi_b(t) = \frac{4\pi d(t)}{\lambda_0}$, where $d(t)$ describes the displacement of the object during the chirp time and $\lambda_0$ is the signal wavelength.

Note that the TI mmWave radar uses a complex band architecture.
It uses a complex mixer (an IQ mixer) to multiply the two signals $S_{tx}$ and $S_{rx}$, which has several advantages like a lower noise figure. 
When in complex form, the IF signal in \Cref{eq:IF1} can be written as:
\begin{equation}\label{eq:IF3}
    IF(t) = B \cdot  e^{j(\omega_{b} t + \phi_{b})}
\end{equation}
which has the same frequency and phase as in \Cref{eq:IF2}.

\section{Steering Vector and AoA Estimation}\label{apx:steering-vector}
\begin{figure}[hb]
    \centering
    \includegraphics[width=0.5\linewidth]{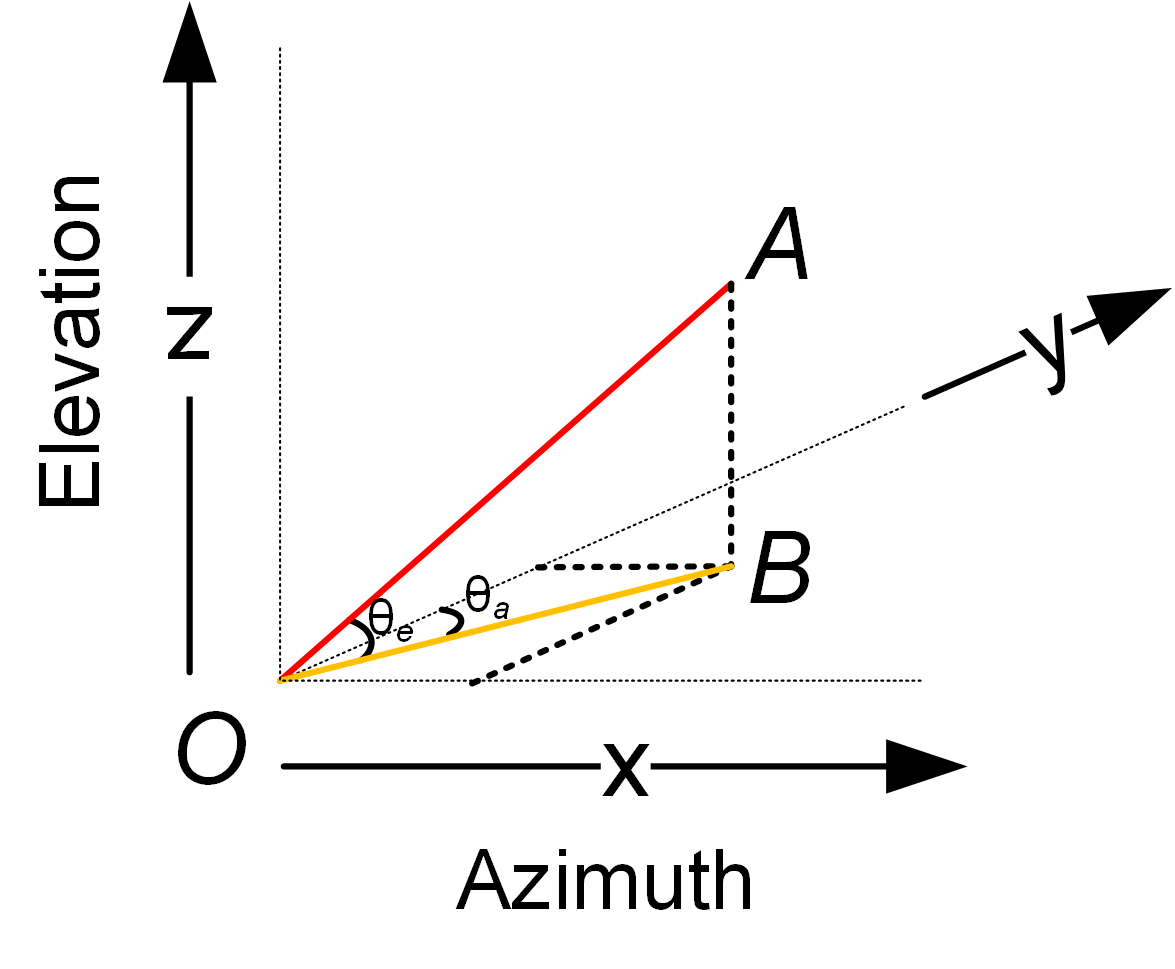}
    \caption{The azimuth and elevation angle of an object. }
    \label{fig:azimuth_elevation}
\end{figure}

\Cref{fig:azimuth_elevation} shows the AoA of an object (point $A$) to the radar (point $O$). 
The azimuth angle $\theta_a$ is defined to be the angle between the object's projection on the horizontal plane and the front direction of the radar.
The line of incidence of the object $OA$ is projected onto the horizontal plane as $OB$, and the angle between $OB$ and the y-axis is the azimuth angle $\theta_a$.
The elevation angle $\theta_e$ is defined to be the angle between the object and the horizontal plane (between line $OA$ and the x-y plane).

The steering vector is a function of the receiver layout and the incident angle.
To introduce the concept of steering vectors, it is easier to start with the one-dimensional situation. 
Assuming there are two receivers separated by $l$, a signal will travel an additional distance $\Delta d$ to reach the second receiver, where the following approximation can be made (as shown in \Cref{fig:angle}):  
\begin{equation}\label{eq:dlsin}
    \Delta d = l\cdot sin(\theta)
\end{equation}
Given that the phase of a sinusoid signal travelled over any distance $\Delta d$ will have a phase $\frac{2\pi \Delta d}{\lambda}$, the phase difference between the two neighbouring receivers will be:
\begin{equation}\label{eq:phase-angle}
    \Delta\phi =2\pi\cdot\frac{\Delta d}{\lambda}=\pi\cdot sin(\theta)
\end{equation}

When using a receiver array with $N$ azimuth receivers,
each subsequent receiver beyond the first one will receive an additional phase change of $\Delta\phi$, 
which can be written as a steering vector:
\begin{equation}\label{eq:steering-vector-1d}
    s(\theta, N) = [1, e^{j\pi\cdot sin(\theta)}, e^{2j\pi\cdot sin(\theta)}, ..., e^{(N-1)j\pi\cdot sin(\theta)}]
\end{equation}

\begin{figure}[hb]
    \centering
    \begin{subfigure}{0.49\linewidth}
        \includegraphics[width=\linewidth]{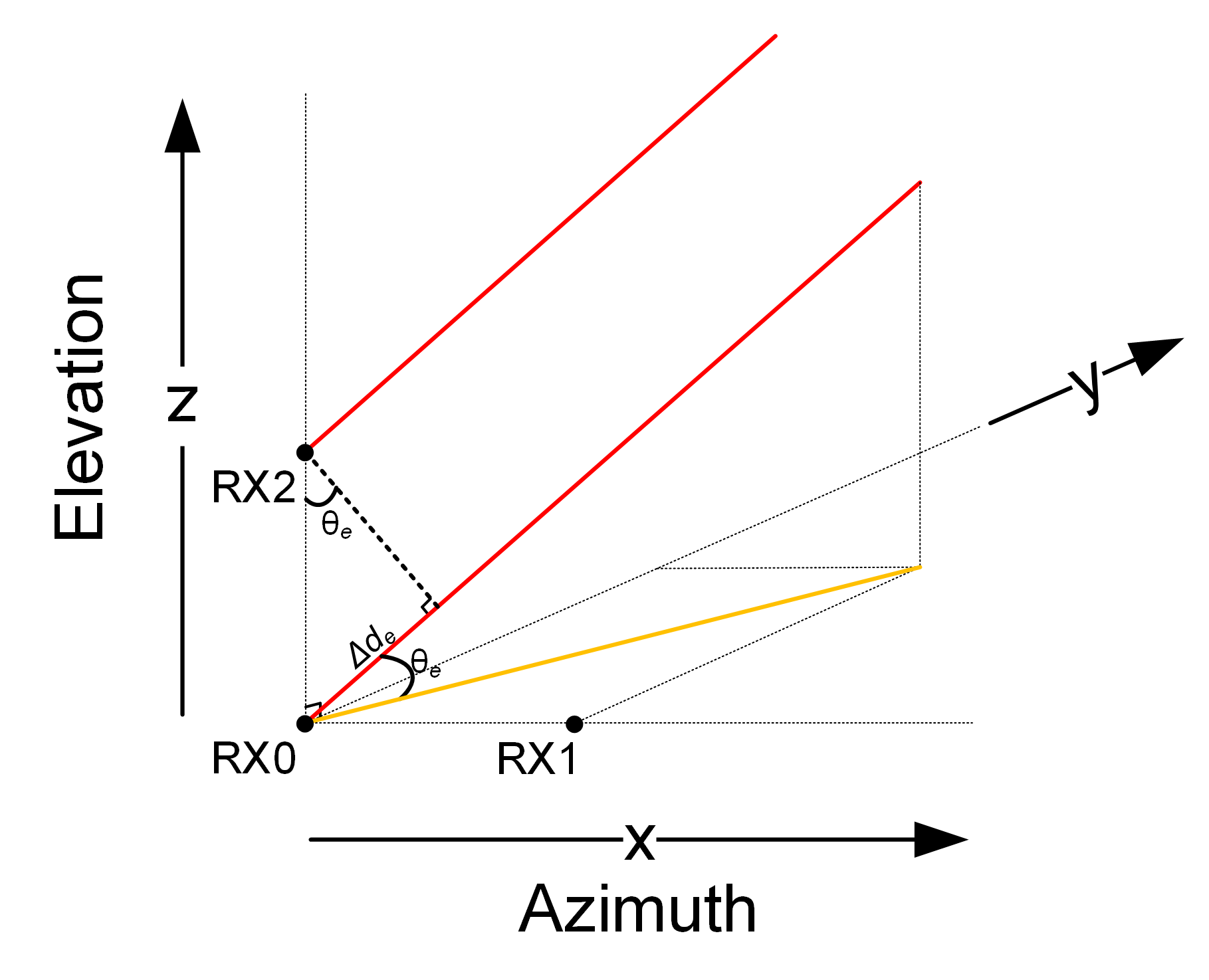}
        \caption{Elevation angle.}
    \end{subfigure}
    \begin{subfigure}{0.49\linewidth}
        \includegraphics[width=\linewidth]{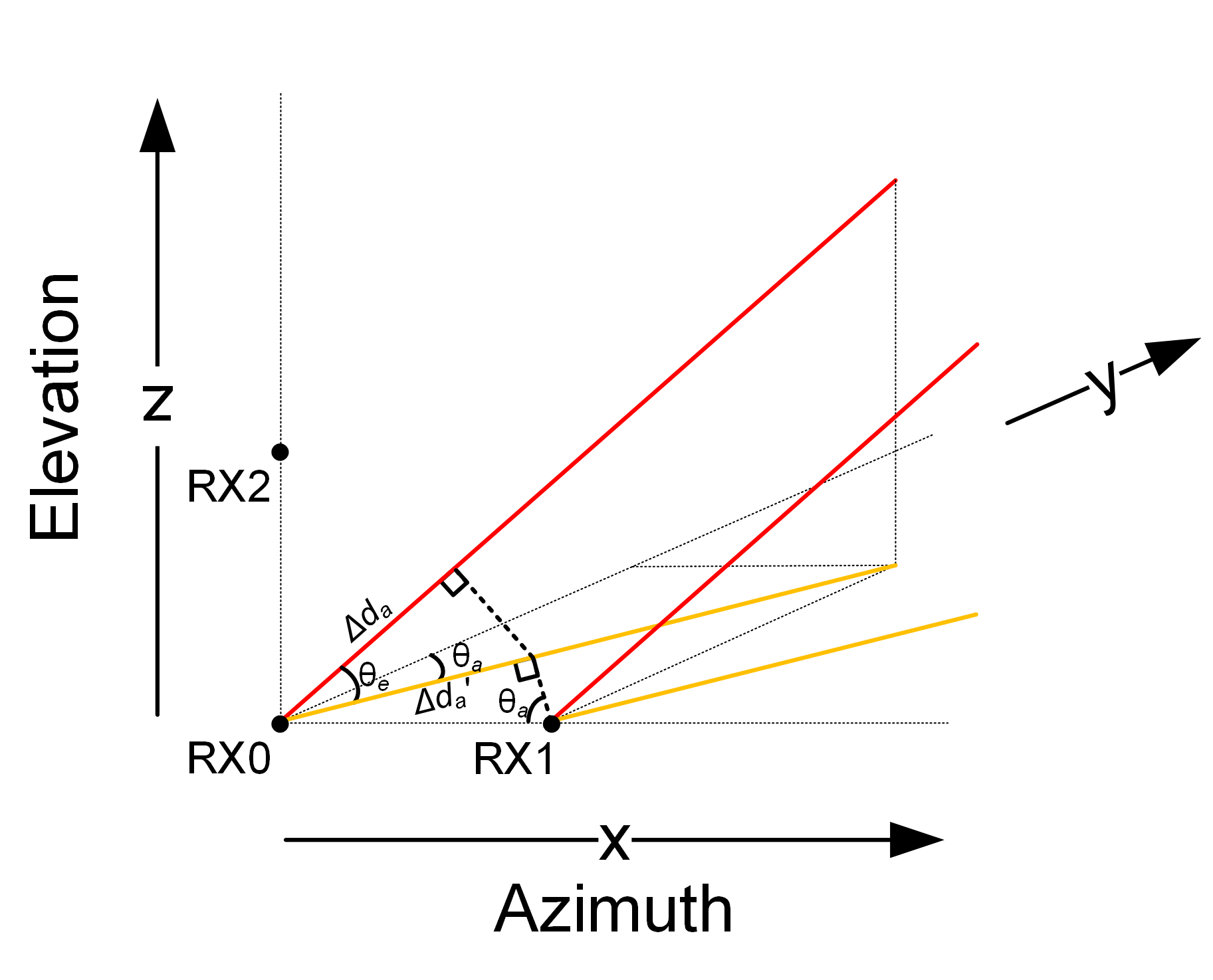}
        \caption{Azimuth angle.}
    \end{subfigure}
    \caption{The AoA can be estimated from the phase difference between adjacent receivers. }
    \label{fig:angles_wxwz}
\end{figure}

When considering the AoA in both azimuth and elevation directions, the situation is shown in \Cref{fig:angles_wxwz}.
Distances $\Delta d_a$ and $\Delta d_e$ represent the extra distance travelled by the signal to reach receiver RX0 when compared with the azimuth receiver RX1 and the elevation receiver RX2 respectively. 
Similar to \Cref{eq:phase-angle}, the estimation of the elevation angle $\theta_e$ is given by: 
\begin{equation}\label{eq:theta_e}
\begin{split}
    sin(\theta_e) &= \frac{\Delta d_e}{l}\\
    &= \frac{\Delta\phi_e}{\pi}
\end{split}
\end{equation}
where $\Delta\phi_e$ is the phase difference between RX0 and RX2. 

The azimuth angle requires a projection from the object's 3D location to the horizontal plane.
As shown in \Cref{fig:angles_wxwz}b, the projection from $\Delta d_a$ to $\Delta d_a\textrm'$ gives:
\begin{equation}\label{eq:projection}
    \Delta d_a=\Delta d_a\textrm'\cdot cos(\theta_e)
\end{equation}
Then, the angle $\theta_a$ can be calculated as:
\begin{equation}\label{eq:theta_a}
\begin{split}
    sin(\theta_a) &= \frac{\Delta d_a\textrm'}{l}\\
    &= \frac{\Delta d_a}{l\cdot cos(\theta_e)}\\
    &= \frac{\Delta\phi_a}{\pi\cdot cos(\theta_e)}
\end{split}
\end{equation}
where $\Delta\phi_a$ is the phase difference between RX0 and RX1.
Extending \Cref{eq:theta_e} and \Cref{eq:theta_a} with multiple receivers as a 2D array gives \Cref{eq:steering-vector-2d}.

Once the values of $\theta_a$ and $\theta_e$ are found, then according to \Cref{fig:angle}, it can be shown that:
\begin{equation}
    \begin{split}
        x &= OB\cdot sin(\theta_a) = OA \cdot cos(\theta_e) sin(\theta_a) = OA \cdot \frac{\Delta\phi_a}{\pi}\\
        z &= OA \cdot sin(\theta_e) = OA\cdot \frac{\Delta\phi_e}{\pi}\\
        y &= \sqrt{OA^2-x^2-z^2}
    \end{split}
\end{equation}
where $OA$ is the distance of the object and can be found from the range-FFT.

\end{document}